\documentclass[final,12pt,authoryear,doi,times]{elsarticle}
\makeatletter
   \def\ps@pprintTitle{%
      \let\@oddhead\@empty
      \let\@evenhead\@empty
         \def\@oddfoot{\centerline{\thepage}}%
      \let\@evenfoot\@oddfoot
   }
\makeatother
\usepackage{lineno}
\modulolinenumbers[5]
\usepackage{amsmath,amssymb,amsthm,amsfonts}
\usepackage{epsfig}
\usepackage{subfigure}
\usepackage{enumerate}
\usepackage{color}
\usepackage{float}
\usepackage{epstopdf}
\usepackage{graphicx}
\theoremstyle{plain}
\usepackage{titleref}
\usepackage{xr}

\usepackage{bm}

\usepackage[utf8]{inputenc}

\begin{document}
\begin{frontmatter}
\title{Nonlinear Inclusion Theory with Application to the Growth and Morphogenesis of a Confined Body}

\author[CEE]{Jian Li\corref{contrib}}
\author[CEE]{Mrityunjay Kothari\corref{contrib}}
\author[AA]{Chockalingam Senthilnathan}
\author[CEE]{Thomas Henzel}
\author[YALE]{Qiuting Zhang}
\author[ME]{Xuanhe Li}
\author[YALE]{Jing Yan\corref{corauthor}}
\author[CEE,ME]{Tal Cohen\corref{corauthor}}

\cortext[corauthor]{Corresponding author}
\cortext[contrib]{These authors contributed equally}
\ead{jing.yan@yale.edu (Jing Yan), talco@mit.edu}

\address[CEE]{Department of Civil and Environmental Engineering, Massachusetts Institute of Technology, Cambridge, MA, USA, 02139}
\address[ME]{Department of Mechanical Engineering, Massachusetts Institute of Technology, Cambridge, MA, USA, 02139}
\address[AA]{Department of Aeronautics and Astronautics, Massachusetts Institute of Technology, Cambridge, MA 02139}
\address[YALE]{Department of Molecular, Cellular and Developmental Biology, Yale University, New Haven, CT 06511, USA.}

\begin{abstract}
One of the most celebrated contributions to the study of the mechanical behavior of materials is due to J.D. Eshelby, who in the late 50s revolutionized our understanding of the elastic stress and strain fields due to an ellipsoidal inclusion/inhomogeneity that undergoes a transformation of shape and size. While Eshelby’s work laid the foundation for significant advancements in various fields, including fracture mechanics, theory of phase transitions, and homogenization methods, its extension into the range of large deformations, and to situations in which the material can actively reorganize in response to the finite transformation strain, is in a nascent state. Beyond the theoretical difficulties imposed by highly nonlinear material response, a major hindrance has been the absence of experimental observations that can elucidate the intricacies that arise in this regime. To address this limitation,  our experimental observations reveal the key morphogenesis steps of \textit{Vibrio cholerae} biofilms embedded in hydrogels, as they grow by four orders of magnitude from their initial size. Using the biofilm growth as a case study, our theoretical model considers various growth scenarios  and employs 
two different and complimentary methods --- a minimal analytical model and finite element computations ---  to obtain approximate equilibrium solutions. A particular emphasis is put  on determining the \textit{natural growth path} of an inclusion that optimizes its shape in response to the confinement, and the onset of damage in the matrix, which together explain the observed behavior of biofilms. Beyond bacterial biofilms, this work sheds light on the  role of mechanics in determining the morphogenesis pathways  of confined growing bodies and thus applies to a broad  range of phenomena  that are ubiquitous in both natural and engineered material systems.  

\end{abstract}


\end{frontmatter}


%


\section{Introduction}\label{intro}

\begin{figure}[H]
  \centering
    \includegraphics[width=0.6\textwidth]{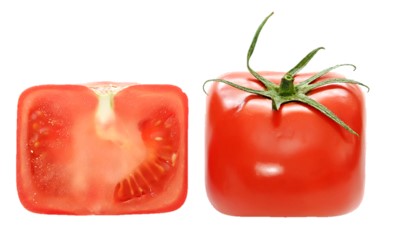}
  \caption{Tomato grown inside a cuboidal box. As the tomato grows, it experiences confinement from the box and adapts its shape resulting in a cuboidal tomato. Once the tomato is fully grown, removing it from the box does not imply recovery of its initially desired shape. Similar process has been applied to other fruits as well to obtain pentagonal oranges and pyramidal watermelons, for instance. Beyond aesthetics, this practice has the potential to substantially reduce storage and transportation costs. (Image credit: Flickr/moonimage)}\label{tomato}
\end{figure}

In nature, morphogenesis is the process by which an organism acquires its shape as it grows. This process results from an interplay between the organism's developmental bluepr  int and environmental factors. The two \textit{driving forces} together determine the fate of the growing body\footnote{Here we distinguish between the terms growth and morphogenesis. The former refers to changes in size  and the latter to changes in shape. }. 
A plant adjusting itself to gather more sunlight (photomorphogenesis) \citep{kendrick2012photomorphogenesis}, adaption of root systems to optimize nutrients and water supply \citep{sutton1980root},  healing of a broken bone in response to  external stresses \citep{nomura2000molecular, ghiasi2017bone}, a tomato achieving a cuboidal shape  when growing under mechanical confinement (Fig. \ref{tomato}) and permanent deformities resulting from ancient foot-binding practices \citep{richardson2009chinese, gu2013deformation} are examples of the strong role that different  local (nutrition, temperature) and global (sunlight, mechanical confinement) environmental factors can play in morphogenesis. 
The intimate coupling between the environment and morphogenesis is further highlighted by noting that the adjustment  of environmental factors does not simply reverse their effect to result in the same final state as the body would have reached had it grown in absence of these factors. 
For instance, removal of mechanical confinement does not return the tomato to a round shape and neither does it return the deformed feet to their regular structure.
It is evident through these examples that the feedback between the environment and morphogenesis continually and progressively determines the shape of the body. However, \textit{how} this crosstalk occurs remains an open question.

\vspace{-5mm}

Beyond nature, processes of growth and morphogenesis are ubiquitous in engineered systems, where understanding the dominant role of environmental factors in determining the evolution of a material is imperative to optimize its durability and sustainability. Precipitation, aggregation, swelling, thermal expansion, phase-transitions, and chemical reactions, can all lead  to internal reorganization in the material  that are akin to the growth and morphogenisis in nature. 
One of the most pressing challenges in this context is   understanding the deterioration of concrete structures due to internal precipitation and swelling \citep{gallyamov2020multi}. Similar processes occur in metals and can alter their mechanical properties \citep{fratzl1999modeling, porter2009phase, ueland2013transition, kothari2019thermo}, and  in advanced fabrication techniques, such as additive manufacturing  and frontal polymerization \citep{robertson2018rapid, goli2019frontal},
where thermal expansion, precipitation and chemical reactions spontaneously emerge in the material. Similarly, biofabrication technologies use living cells and extracelluar matrix to construct tissue-like structures \citep{morley2019quantitative}. However, understanding how these constructs continue to adapt and change their morphology after fabrication  is imperative to tailor the resulting tissue as well as its  desired functionalities \citep{deo2020bioprinting,levato2020shape}.

In all of the examples mentioned above, the growing body can be highly heterogeneous with regions that may  evolve over time through various mechanisms and can lead to large deformations and failure. In this work, we limit our attention to what can be considered as an elemental unit within the larger body, a growing and morphing \textit{inclusion} embedded in an unbonded homogeneous medium. We  focus on the role of the mechanical confinement on the morphogenesis.

Growth of a single  body in a mechanically confined environment is ubiquitous not only as a representative unit within a heterogeneous medium, but also in diverse examples ranging across length scales. Embryogenesis is perhaps the most fundamental example where the extreme growth of an embryo leads to large deformation  of the embedding medium. Other examples  include growth of biomolecular condensates inside cells \citep{banani2017biomolecular}, growth of tumors inside normal tissues \citep{balkwill2012tumor}, and  growth of plant roots in soil.
Liquid-liquid phase separation, if it occurs within a solid matrix, provides another example where the nonlinear effects are significant \citep{rosowski2020elastic, kothari2020effect, kim2020extreme, wei2020modeling, ronceray2021liquid}. The organization and dynamics of the separating phases, as they condense and grow, is strongly influenced by the elastic resistance of the matrix. 
Finally, emerging techniques for characterization of soft materials grow fluid filled cavities inside the material to estimate its nonlinear properties \citep{kundu2009cavitation, raayai2019intimate, yang2019hydraulic, chockalingam2021probing}. Notably, these experiments also indicate morphological transitions from regular-shaped cavities to  branched fracture patterns as the cavity grows \citep{raayai2019intimate,yang2019hydraulic,morelle2021visualization}.

Any discussion on the mechanics of an embedded body growing inside another body is incomplete without acknowledging the pioneering work of  \cite{Eshelby1957TheDO, eshelby1959elastic}.
Through an elegant sequence of `imaginary cutting, straining and welding operations' Eshelby calculated the elastic fields (strain, stress and displacement) in an infinite medium (the matrix) that contains an embedded ellipsoidal region (the inclusion) undergoing stress-free strain (the transformation strain) and obtained closed-form analytical solutions for special cases \citep{Eshelby1957TheDO}.
This  work laid the foundation for significant advances in mechanics, materials science, and geomechanics \citep{lee1977elastic, johnson1988precipitate, fratzl1999modeling, khachaturyan2013theory, meng2014eshelby, krichen2019liquid, sharma2004size}. However, Eshelby's solutions are limited to linear elasticity and therefore apply  only at the limit where changes of volume and shape are infinitesimal. 

To model morphogenesis one must forfeit the luxuries of linear elasticity which enabled `imaginary cutting, straining and welding' operations by virtue of the superposition principle \citep{diani2000problem, yavari2013nonlinear}.
Moreover, systems undergoing large strains are inevitably prone to damage. This presents a unique challenge from the modeling point of view as these different physical scenarios translate into an evolving reconfiguration of both the inclusion and the matrix\footnote{
We note that in the classical literature \citep{Eshelby1957TheDO, mura2013micromechanics} the term `inclusion' is used to describe the embedded region when it has the same elastic properties as that of the matrix. The term `inhomogeneity' is employed when the embedded region differs in elastic properties from the medium. While we are cognizant of this terminology, in the regime of nonlinear constitutive response, both the embedded region and the medium can exhibit different properties depending on their local deformation state and therefore the distinction based on material properties does not carry over from the linear theory. In this work, we  use the term inclusion more freely for any embedded body undergoing growth and morphogenesis.}.
In this context, the current work will address two major challenges and gaps in the existing literature. 
Firstly, to the best of our knowledge, the pivotal question of determining the \textit{natural} growth trajectory of a growing material system remains unanswered\footnote{In a continuum formulation this translates to determining the natural  evolution of the growth tensor, as opposed to prescribing it through a kinematics motivated constitutive law.}. 
This motivation for a natural growth law was concisely stated by D'arcy Thompson in his seminal work \citep{thompson1942growth}: \textit{``An organism is so complex a thing, and growth so complex a phenomenon, that for growth to be so uniform and constant in all the parts as to keep the whole shape unchanged would indeed be an unlikely and an unusual circumstance. Rates vary, proportions change, and the whole configuration alters accordingly."} 
In this spirit, this work takes an alternative view on the question of growth and morphogenesis in an attempt to understand how
the configuration alters in response to external mechanical constraints.  
Secondly, obtaining analytical solutions for growth problems has been challenging due to the ensuing large deformations and nonlinear constitutive response. Existing solutions in the literature are thus limited to simple geometries.  Here we develop an approximate quasi-analytical theory that can capture the nonlinear morphogenesis of a growing inclusion and we  compare it with finite element simulations.

Next, before proceeding to describe our theory (Sections \ref{sec:def} and \ref{sec:equil}) and its results (Section \ref{sec:results}),  we further establish the notion of a natural growth path and the possible emergence of various growth scenarios in nonlinear deformation through observations of the growth and morphogenesis of confined biofilms.

\section{Case Study: Confined growth of Bacterial Biofilms}\label{sec:CS}
In this work, we develop a  theory that  can apply to  material systems in which the growing body is not intrinsically programmed to follow a particular blueprint, and as such its growth path depends solely on the external constraints. At first, such an assumption on the growth scenario may seem limiting, but in fact its manifestations are ubiquitous in both natural and engineered systems. A quintessential example is the growth of bacterial biofilms \citep{hall2004bacterial, mukherjee2019bacterial, dufrene2020mechanomicrobiology}. These biofilms are aggregates of cells that duplicate at a steady time period, given enough nutrition and ideal conditions, leading to exponential growth in time.
With the extracellular matrix serving as a glue that holds the cells together, the resulting macroscopic formation behaves like a soft solid with measurable mechanical properties (e.g. mechanical stiffness, viscoelasticity, and yield stress) \cite{yan2019mechanical}. In absence of mechanical confinement, biofilm organization is random; the biofilm cluster takes an irregular shape when suspended, and hemispherical when attached to a solid substrate \citep{yan2016vibrio}.

Biofilms are present almost everywhere on our planet; they perform several functions that are an essential part of carbon turnover in the environment \citep{ebrahimi2019cooperation, enke2018microscale}, but can also become a nuisance that leads to fouling of ships \citep{de2018marine}, or  water filtration systems \citep{liu2016understanding}. Biofilms also grow in our bodies -- pores in our skin  \citep{alexeyev2013bacterial}, teeth \citep{saini2011biofilm}, and gut \citep{probert2002bacterial} are  examples of hot spots for bacterial colonies, which can also grow on surgical implants, and in chronic infections \citep{rybtke2015pseudomonas}. Forming biofilms drastically enhances the resistance of the individual cells to antibiotics,  and thus accounts for a significant part of all human microbial infections \citep{bryers2008medical}.  Understanding the growth of biofilms in settings that simulate their natural environment is thus essential in the pursuit of future therapies \citep{vasudevan2014biofilms}. Beyond medicine, it can elucidate various processes in the environment and can help to alleviate their unwanted growth in engineered systems.

While several studies have examined the formation of biofilms on flat substrates \citep{garrett2008bacterial, seminara2012osmotic, fei2020nonuniform, song2015effects}, it is only recently that their growth and morphogenesis in confined three-dimensional settings has been observed. This has been made possible due to the development of cell-level imaging techniques \citep{rachel} that track the formation of  bacterial colonies starting from a single seed bacteria embedded in the bulk of a confining medium. Here, we use the same experimental system to further examine the sensitivities of the growth process in comparison with the theoretical predictions. In particular,  we focus on the influence of  material properties on the morphogenesis. 

\begin{figure}[H]
 \center
 \includegraphics[width=\textwidth]{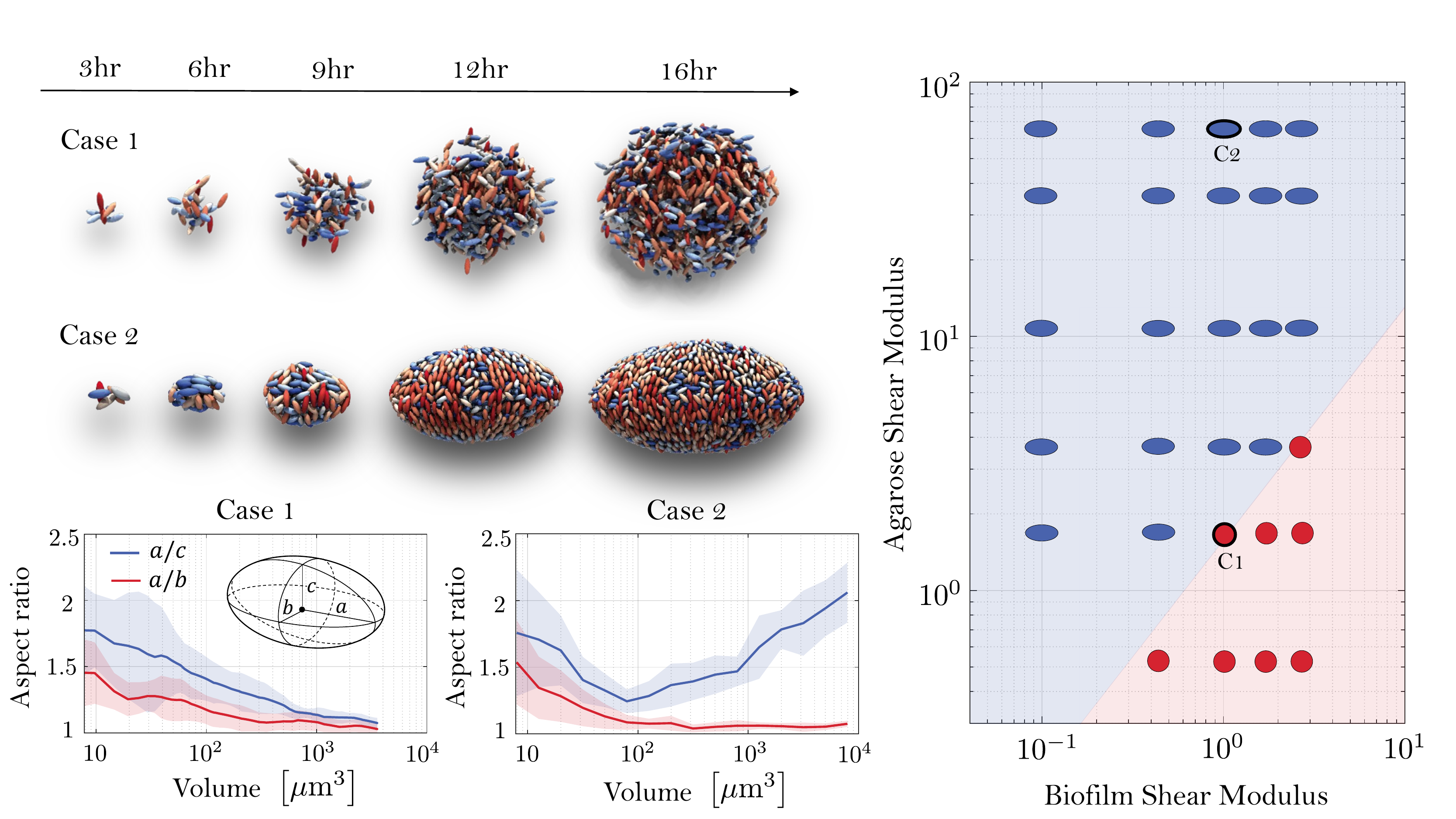}
 \caption{Bacterial Biofilm Observations. (top left) Time-lapse images of biofilms reconstructed in 3D at the level of a single cell. Color coding is defined by the orientation of the cells in respect to the minor axis.   Two different cases are shown  and their corresponding moduli are indicated in the phase diagram. The length of a single bacteria is $\sim 1\mu \text{m}$; (bottom left) evolution of the biofilm aspect ratios for the two cases. For Case 1 - behavior is monotonic. For Case 2 - transition towards an oblate shape is observed; (right) phase diagram distinguishing two typical morphogenesis paths, i.e monotonic (red) and non-monotonic(blue) as shown for Cases 1 and 2, and estimated based on combination of time evolution data, and final aspect ratios of mature colonies. Additional experimental details can be found in \cite{rachel}.}
 \label{CaseStudy}
 \end{figure}

In our experimental system, isolated  \textit{Vibrio cholerae} bacteria are embedded inside an agarose gel where they begin to multiply to form a biofilm. There are three key features of this system  that are imperative to enable interpretation of the mechanical phenomena:  
\begin{itemize}
    \item[\textit{(i)}] \textit{Biofilm morphogenesis is captured with high spatiotemporal resolution over the full course of its growth.} This is achieved using an adaptive imaging technique \cite{yan2016vibrio} that enables visualization of both the global morphology and the single-cell level architecture  of an embedded biofilm containing up to $10^4$ cells. The experimental technique allows us to fully reconstruct the biofilm morphology and its internal architecture  \textit{in silico}  at different times throughout its growth, as shown for two example cases in Fig. \ref{CaseStudy}. This also confirms that there is no penetration of single bacteria into the confining matrix, and that there is no fracturing.  
    \item[\textit{(ii)}] \textit{The stiffness of the confining medium and the biofilm are separately tunable.} This is an important and unique feature of this system that enables us to determine the constitutive sensitivity of morphogenesis over a broad range of properties. The stiffness of the biofilm is tuned by mutagenesis. Five different mutants that lack one or more extracellular matrix components are used \citep{yan2018bacterial} and correspond to biofilm shear moduli in the range $\sim 0.1-10$ kPa. The stiffness of the confining body is tuned by adjusting the concentration of the agarose matrix. Varying the concentration in the range $0.3-2\%$  translates to shear moduli in the range $\sim 1-100$ kPa. 
    \item[\textit{(iii)}] \textit{Bacteria reproduce at a constant rate that is not influenced by the confining medium.} The agarose gel matrix is biocompatible and   infused with sufficient nutrient to sustain the biofilm growth. Additionally, we use the well characterized rugose \textit{Vibrio cholerae} strain that is locked in a high cyclic diguanylate level \citep{beyhan2007smooth}. This insensitivity is confirmed by counting the number of cells as a function of time for different biofilm/agarose combinations and observing exponential growth with a division time that is independent of the external confinement. The constant reproduction rate also confirms that the local growth rate of the biofilm is uniform and constant throughout the process. 
\end{itemize}
For more details on the experimental protocols and measurement techniques, the reader is referred to \cite{rachel}.

The experimental results in Fig. \ref{CaseStudy} show that the confinement plays a crucial role in driving the morphogenesis of biofilms. First, we focus on two example cases (shown on the left side of the figure). The morphology of the observed ellipsoid-like shapes can be quantified  by two aspect ratios; $a/c$ - the ratio between the  major and minor axes, and $a/b$ - the ratio between the major and median axes  (see inset).  Although the initial shape (with only few bacteria) is not well defined, after  $\sim 6$ hours a clear shape emerges and begins to evolve. 
The two example cases exhibit distinct morphogenesis: In both cases we initially observe a gradual tendency towards a spherical shape, with both $a/c\to1$ and $a/b\to1$. This tendency persists in Case 1, whereas in Case 2 a   shift is observed at $ V\sim100\mu\rm{m}^3$; the tendency of $a/c$ flips and the biofilm morphology tends towards an oblate spheroid shape. The phase diagram in Fig. \ref{CaseStudy} examines the influence of the respective stiffnesses of the biofilm and the agarose, giving rise to these two typical morphological trends. It is shown that the transition towards an oblate shape emerges if the agarose is stiffer than the biofilm. Otherwise, the tendency towards a spherical shape is maintained. 

In this work we aim to explain the emergence of these different trends as well as their constitutive sensitivity.  Although it has been experimentally confirmed that fracture is not culpable at this scale \citep{rachel,kim2020extreme},  arge deformations induced by a body that has grown by four orders of magnitude must be accompanied by damage, which will be considered in this work. Additionally, our models will allow us to decipher between the separate roles of deformation and growth in determining the observed shapes\footnote{In \cite{rachel} it was shown that upon dissolving the agarose around a mature oblate  biofilm, its recovered stress free aspect ratio $a/c$ further increases.}.   

\section{Problem definition and modeling approach}\label{sec:def}

\begin{figure}[h]
 \center
 \includegraphics[width=0.8\textwidth]{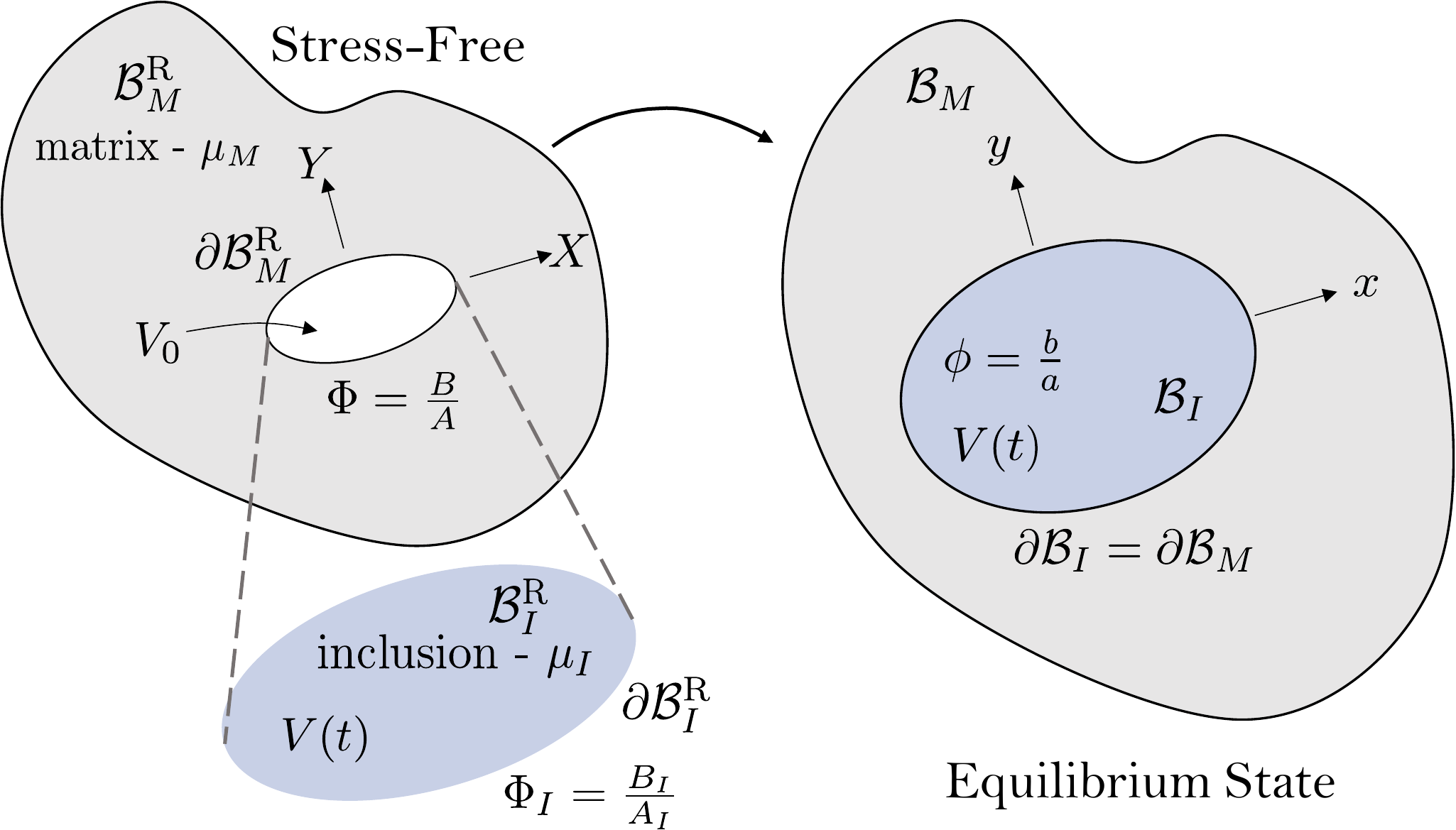}
 \caption{Schematic illustration of the stress-free and deformed states of the inclusion-matrix system.}
 \label{illus}
 \end{figure}

Consider a body that can change its shape as it grows. This body, the \textit{inclusion}, is embedded within an unbounded confining medium, the \textit{matrix}. 
In its undeformed, stress-free state, the matrix occupies the region $\mathcal{B}_M^{\rm R}$. 
 This state can be achieved if the grown inclusion is removed from the matrix (Fig. \ref{illus}). 
 The boundary of the remaining \textit{void} is denoted by $\partial\mathcal{B}_M^{\rm R}$; its undeformed volume is denoted by $V_0$. 
 The volume of the inclusion, $V(t)$, varies with time\footnote{Here `time' (denoted by $t$) is a surrogate variable that determines the process. 
 No dynamic, or rate dependent effects are included in this work.}; we assume that if it were removed from the matrix, it would recover an undeformed stress-free state\footnote{We restrict our attention to compatible growth of the inclusion. This simplification allows us to focus on   incompatibility between the inclusion and the matrix.  }  that occupies the region $\mathcal{B}_I^{\rm R}(t)$, and its boundary is denoted by $\partial\mathcal{B}_I^{\rm R}(t)$. 
\textit{In their stress-free states, the inclusion and the matrix are incompatible. }
In the physical space, the matrix and the inclusion occupy the regions  $\mathcal{B}_M(t)$ and  $\mathcal{B}_I(t)$, respectively, and their boundaries  intersect, i.e. $\partial \mathcal{B}_M(t)=\partial \mathcal{B}_I(t)$, to enclose the volume $V(t)$.

\bigskip

\noindent\textbf{Deformation gradient.} At a given time, in the undeformed state, material points are labeled using the Cartesian coordinate system $\bm{X}=(X,Y,Z)$ where $\bm X\in \mathcal{B}^{\rm R}_M(t)$ or $\bm X\in \mathcal{B}^{\rm R}_I(t)$, for the matrix and the inclusion, respectively. Upon deformation, a placement map $\bm{x}(t)={\bm{\chi}}({\bm{X}},t)$ assigns the material points their location 
in the physical space, as labeled by the Cartesian  coordinates $\bm{x}(t)=(x(t),y(t),z(t))$ where $\bm x\in \mathcal{B}_M(t)$ or $\bm x\in \mathcal{B}_I(t)$, respectively. Accordingly, we can write the deformation gradient $\mathbf{F}(t)=\partial \bm{\chi}/\partial \bm{X}$. Note that this mapping is defined from the stress-free state at time $t$. 
\bigskip

\noindent\textbf{Configurational tensor.} In this work we account for changes that may occur in the configurations (i.e. the stress-free states) of both the inclusion and the matrix, which can be affected, for example, by growth, remodeling, and damage. The mathematical treatment that we employ transcends different \textit{configurational forces}, hence we coin the general term - \textit{configurational tensor}.  
If one can identify an initial state of the body, at say $t=0$, such that $\bm{X}_0=\bm{X}(0)$, then a placement map $\bm{X}(t)={\bm{\chi}_0}({\bm{X}}_0,t)$ can be defined and the configurational tensor is $\mathbf{F}_0(t)=\partial \bm{\chi}_0/\partial \bm{X}_0$. If the changes in configuration are due to growth, then $\mathbf{F}_0(t)$ is more commonly referred to as the \textit{growth tensor}\footnote{The common multiplicative form of the deformation gradient (from an arbitrary initial state ${\bm{X}}_0$) can be written as $\hat{\mathbf{F}}(t)=\mathbf{F}\mathbf{F}_0$, and follows from the differentiation of $\bm{x}(t)={\bm{\chi}}({\bm{\chi}_0}({\bm{X}}_0),t)$ by making use of the chain rule.}. 

\bigskip
\noindent\textbf{Remote field.} In the remote field we require that the influence of the growing inclusion vanishes, and thus we apply the condition
\begin{equation}\label{RC0}
\mathbf{F}\to \mathbf{F}_\infty  \quad \text{for} \quad |\bm{x}|\to\infty, 
\end{equation}
where $\mathbf{F}_\infty$ is an applied deformation.

\bigskip
\noindent\textbf{The interface.} It is instructive to parameterize the collection of material points on the boundaries  of the matrix and the inclusion (at a given time) by\footnote{Note that from hereon the subscript $(\cdot)_I$ will denote values for the inclusion. Also, for mathematical compactness, we omit the time dependence, nonetheless all fields and boundaries in both the reference and current frames can vary with time.}  \begin{equation}
    \bm{X}_b=\bm{X}_b(s) ~\in~ \partial\mathcal{B}_M^{\rm R} \quad \text{and} \quad \bm{X}_{bI}=\bm{X}_{bI}(s_I)~\in~ \partial\mathcal{B}_I^{\rm R},
\end{equation} respectively\footnote{Note that $s\in\mathbb{R}^1$ in a 2D setting, and $s\in\mathbb{R}^2$ in a 3D setting.}. Upon deformation, compatibility implies that these boundaries map to the same surface, namely
\begin{equation}
    \bm{x}_b(s)=\bm{\chi}(\bm{X}_b(s)), \quad \bm{x}_{bI}(s_I)=\bm{\chi}_I(\bm{X}_{bI}(s_I)) ~~\in~~  \partial \mathcal{B}_M=\partial \mathcal{B}_I.
\end{equation}
Mathematically this implies that there exists a transformation $s_I=\xi(s)$ such that  \begin{equation}\label{compt}
   \bm{x}_b(s)=\bm{x}_{bI}(\xi({s})). 
\end{equation}
Namely, if the interface is free to slip, $\xi(s)$ is a degree of freedom in the system, whereas if the interface is strictly adhered, such that there is correspondence between points on either side of the interface,  then  $\xi(s)$ is a prescribed kinematic constraint. In this work we will consider both limits of strictly bonded and free to slip.  We will further discuss the consequences of this interface constraint in the following sections.


\noindent With these definitions in place we can now state the primary question that we seek to answer: 

\begin{center}
 {\bf{Question A:}}   \textit{Given the volume $V(t)$ of an inclusion, what shape, $\bm{X}_{bI}$, would it take?}
\end{center}

\smallskip 

This question centers on determining changes in \textit{configuration},  which may ultimately affect changes in the observed deformed morphology. Changes in  configuration may follow an inherent path that is encoded in the growing system,  or may arise solely as a response to the mechanical  constraint. In what follows we will further refine this problem statement as it applies to  different possible growth scenarios and may be affected by damage in the matrix. 

Upon obtaining an answer to our primary question, an additional  question that we seek to answer is a nonlinear analogue to the classical Eshelby problem:

\begin{center}
 {\bf{Question B: }}\textit{Given the volume $V(t)$ of an inclusion and its  undeformed shape, $\bm{X}_{bI}$, what deformed shape, $\bm{x}_{b}$, would we observe in the physical space?}

\end{center}

\smallskip

For simplicity, we will restrict our attention to incompressible materials under plane-strain conditions, such that ${\rm det}\mathbf{F}=1$ and $z\equiv Z$.
Additionally, in what follows,  we neglect effects of interfacial friction, surface tension, and rate dependence\footnote{In certain material systems and at different scales these effects may become important and should be subject for future work.} and we model the constitutive behavior of both bodies via  free energy density that depends only on the deformation gradient $\mathbf F$.

\bigskip

\noindent \textbf{  \textit{Scenario \#1: General anisotropic growth}}

\begin{figure}[h]
  \makebox[\textwidth][c]{\includegraphics[width=1\textwidth]{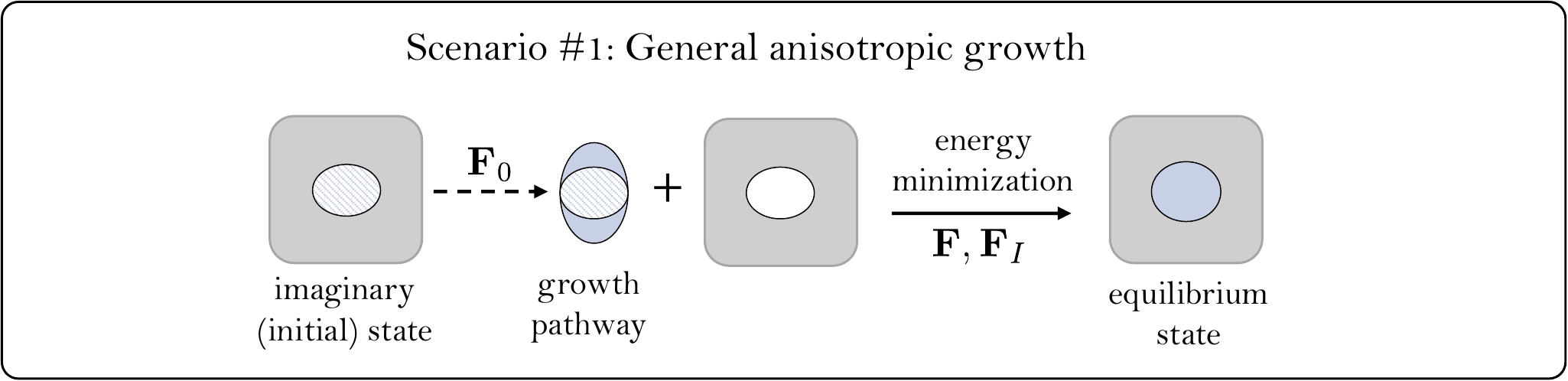}}%
  \caption{Schematic of Scenario \#1.}
\label{fig:S1}\end{figure}

 We begin by considering growth scenarios that are kinematically prescribed.  Namely, at a given time, $t$, with corresponding $V(t)$, the  stress-free configurations of both the inclusion and the body (i.e. $\bm{X}_b$ and $\bm{X}_{bI}$) are known, as illustrated schematically in Fig. \ref{fig:S1}. Such situations may arise, for example, if the growth is a result of thermal expansion or is biologically encoded in the inclusion. With the kinematics fully defined, we have a trivial answer to Question A, and it remains to determine the deformed state of the bodies to answer Question B. 
 Hence, we now proceed to consider the potential energy of the two-body system to examine possible solutions that minimize this energy. 
 
 Let $\mu_M\Psi(\mathbf F)$ be the strain energy density per unit volume, ${\rm dV}$, of the matrix, and let $\mu_{I} \Psi_{I}({\mathbf F}_I)$ be the strain energy density per unit volume, ${\rm dV}_I$, of the inclusion, where the coefficients $\mu_M$ and $\mu_{I}$ are the  shear moduli of the matrix and the inclusion in the linear range, respectively. The total potential energy in the system is 
\begin{equation}\label{S0}
\mathcal{L}(\mathbf{\chi},\mathbf{\chi}_I,{\mathbf F},{\mathbf F}_I,\xi,p)=\mu_M\int\displaylimits_{\mathcal{B}_M^{\rm R}}\Psi(\mathbf F){\rm d} {\rm V}+\mu_I\int\displaylimits_{\mathcal{B}_I^{\rm R}}\Psi_{I}(\mathbf{F}_I){\rm d}{{\rm V}_I}+p[\bm{x}_b(s)-\bm{x}_{bI}(\xi({s}))], 
\end{equation} 
where the last term arises as a response to the compatibility constraint \eqref{compt} and $p$ is a Lagrangian multiplier. 




For the general case (with arbitrary boundary geometries)  an equilibrium solution can be obtained numerically by minimizing the above functional. This can be particularly challenging as the interface condition must be applied with respect to the deformed configuration. Alternatively, if the deformed shape of the interface $\bm{x}_b$ and the corresponding parameterizations $(\xi)$, are prescribed, the energy functional simplifies to the form
\begin{equation}\label{S00}
\mathcal{L}(\mathbf{\chi},\mathbf{\chi}_I,{\mathbf F},{\mathbf F}_I;\bm{x}_b)=\mu_M\int\displaylimits_{\mathcal{B}_M^{\rm R}}\Psi(\mathbf F){\rm d} {\rm V}+\mu_I\int\displaylimits_{\mathcal{B}_I^{\rm R}}\Psi_{I}(\mathbf{F}_I){\rm d}{{\rm V}_I}. 
\end{equation} 
Among all possible deformed shapes, $\bm{x}_b$, for a given reference configuration defined by the set $(V_0, V(t), \bm{X}_b,\bm{X}_{bI})$,  an equilibrium solution would select the shape that minimizes the energy, namely 
\begin{equation} \label{axiomB}
    \bm{x}_b= {\rm arg} ~  \underset{\bm{\bar{x}_b}}{\rm min} ~\left\{{\rm min} ~    \mathcal{L}(\mathbf{\chi},\mathbf{\chi}_I,{\mathbf F},{\mathbf F}_I;\bm{\bar{x}}_b)\right\}.
\end{equation}
The energy minimization in \eqref{axiomB} provides the formal solution to Question B. Upon substituting $\bm{x}_b$ and the corresponding placement maps back in \eqref{S0}, we can write the minimal energy for a given set of parameters as 
\begin{equation}\label{Lmin}
    \mathcal{L}_{min}=\mathcal{L}_{min}(V_0, V(t), \bm{X}_b,\bm{X}_{bI}),
\end{equation}
and the  deformed shape as 
\begin{equation}\label{xb}
    \bm{x}_b=\bm{x}_b(V_0, V(t), \bm{X}_b,\bm{X}_{bI}).
\end{equation}

 In this work we will use computational simulations and  quasi-analytical  methods to obtain solutions for arbitrary anisotropic growth scenarios (as detailed in Section \ref{sec:equil}). These solutions, and their corresponding  energy \eqref{S0} will serve as a foundation to investigate more complex growth scenarios. 
 
\bigskip
 
 \noindent\textbf{Elliptic stress-free configurations.} While any stress-free configuration can be considered, we will limit our attention to an elliptic family of stress-free configurations. Hence, the shapes of the void and the inclusion are uniquely prescribed by their aspect ratios $\Phi=B/A$ and $\Phi_I=B_I/A_I$,  with $A,A_I$ the semi-minor axes, and $B,B_I$ the semi-major axes of the void and inclusion, respectively (Fig. \ref{illus}).  In the deformed state, although the shape may not be of an ellipse, we define a corresponding aspect ratio $\phi=b/a$. Accordingly, we can rewrite the energy minimization in \eqref{axiomB}, such that among all possible deformed aspect ratios, $\phi$, for a given reference configuration defined by the set $(V_0, V(t), \Phi,\Phi_I)$,  an equilibrium solution would select the shape that minimizes the energy, namely 
\begin{equation} \label{axiomB_el}
    \phi= {\rm arg} ~  \underset{{\bar \phi}}{\rm min} ~\left\{{\rm min} ~    \mathcal{L}(\mathbf{\chi},\mathbf{\chi}_I,{\mathbf F},{\mathbf F}_I;\bar{\phi})\right\}.
\end{equation}
The corresponding energy, and deformed aspect ratio then simplify to the respective  forms
\begin{equation}\label{Lmin_ell}
    \mathcal{L}_{min}=\mathcal{L}_{min}(V_0, V(t), \Phi,\Phi_I),\qquad     \phi=\phi(V_0, V(t), \Phi,\Phi_I).
\end{equation}

\noindent \textbf{  \textit{Scenario \#2: An energy-based growth law}}

Here we consider growth scenarios that lack any developmental blueprint to guide the growth, such as the bacterial biofilms described in Section \ref{sec:CS}. In the absence of such a blueprint, mechanical confinement plays the decisive role in determining the morphology of the system and the answer to Question A is no longer trivial. 
\smallskip

\textit {We postulate that, in such growth scenarios, morphogenesis will proceed in a way to minimize the total mechanical energy of the system}.

\smallskip

\begin{figure}[h]
  \makebox[\textwidth][c]{\includegraphics[width=1\textwidth]{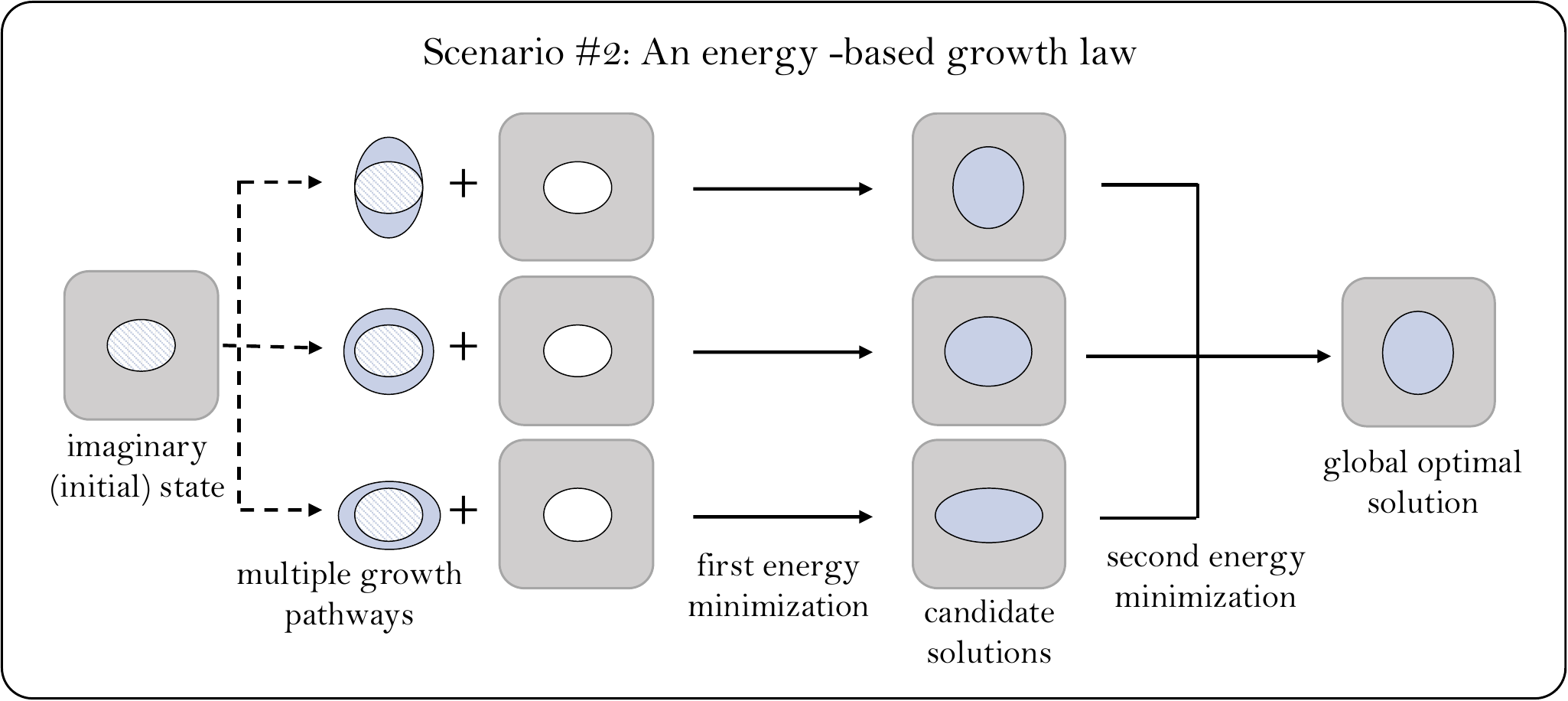}}%
  \caption{Schematic of Scenario \#2.}\label{fig:S2}
\end{figure}

 This postulate introduces the stress-free shape of the inclusion, $\bm{X}_{bI}$, as an additional degree of freedom in the system, thus implying that the inclusion can explore multiple growth pathways (see  illustration in Fig. \ref{illus}). For each growth pathway $\bm{X}_{bI}$ is known, hence the answer to Question B follows directly from \eqref{axiomB}, and provides a sequence of \textit{candidate solutions}, $\bm{x}_{b}=\bm{x}_{b}(\bm{X}_{bI})$, that are in mechanical equilibrium \eqref{xb}. Question A is subsequently answered by selecting  among all candidate solutions  the one with least energy to obtain the optimal shape $\bm{X}_{bI}$, as illustrated schematically in Fig. \ref{fig:S2}. 

By building on the machinery developed in  Scenario \#1, we can write the above solution procedure as
\begin{equation} \label{axiomA}
    \bm{X}_{bI}= {\rm arg} ~  \underset{\bm{\bar X}_{bI}}{\rm min} ~\left\{{\rm min} ~    \mathcal{L}(\mathbf{\chi},\mathbf{\chi}_I,{\mathbf F},{\mathbf F}_I;\bm{x}_b({\bm{\bar{X}}_{bI}}))\right\}.
\end{equation}
The above energy minimization provides the formal solution to Question A for Scenario \#2. Specializing this solution  procedure to the case of elliptic stress-free configurations, reads
\begin{equation} \label{axiomA_el}
    \Phi_I= {\rm arg} ~  \underset{\bar{\Phi}_I}{\rm min} ~\left\{{\rm min} ~    \mathcal{L}(\mathbf{\chi},\mathbf{\chi}_I,{\mathbf F},{\mathbf F}_I;{ \phi}({\bar{\Phi}_I}))\right\}.
\end{equation}

In this scenario we have assumed that the confining body accommodates the growth and reconfiguration of the inclusion through deformation. Since at every stage of growth the system optimizes over its stress-free shapes to select the candidate with the least total energy, the global optimal solution is path independent (Fig. \ref{fig:S2}). Next, we will consider two scenarios by which the matrix can also evolve its configuration to accommodate the growth.  


\bigskip
\noindent \textbf{  \textit{Scenario \#3: Damage induced shape change}}\label{scenario3}

The large deformations experienced by the confining matrix as the inclusion expands its volume can induce damage. For example, in our Case Study (Section \ref{sec:CS}) the growth of an inclusion, from a single bacteria to a colony of $10^4$ cells, inevitably introduces extreme deformations that cannot be elastically sustained by the agarose gel. While different failure mechanisms may emerge, here we will focus our attention to diffuse damage, which is  the primary mode at small-scales \citep{rachel,kim2020extreme}, before fracture initiates. Various mechanisms can be involved in damage processes; they depend on the specific material system, and can be modeled by increasingly complex methods \citep{chocka_zhang2010review,chocka_besson2010continuum,chocka_ambati2015review,chocka_de2002fracture, chocka_narayan2021fracture, chocka_talamini2018progressive, chocka_wu2016stochastic, chocka_miehe2014phase,chocka_raina2016phase,chocka_keralavarma2016criterion}  which include various model parameters.  Without a particular material or damage mechanism at hand, in what follows, we will establish a minimal approach to capture the influence of damage on the observed shape of the inclusion. 

Within the term `damage' we distinguish between irreversible deformation (that does not incur shape change), and volume preserving changes in the matrix configuration. This distinction acknowledges the fact that inelastic deformation does not necessarily lead to changes in shape\footnote{As for example in spherically symmetric expansion of a cavity.  }.
At onset of damage, without unloading, the energetic cost of irreversible deformations is readily captured by the general constitutive model $\Psi(\mathbf{F})$, while changes in configuration  can be interpreted as a change in the \textit{effective} stress-free shape of the void\footnote{Here the term `effective' indicates that this shape may not be the actual observed shape, since damaged material is not removed.}. The change in configuration is associated with a configurational tensor, $\mathbf{F}_0$, which transforms $\bm{X}_b$ to a   new effective shape by $\bm{X}^d_{b}$. 
This transformation also  incurs an energetic cost. If we can estimate the volume of material that has undergone shape change, say $V_d=V_d(\mathbf{F}_0)$, we can assign to it an energy cost
\begin{equation}\label{Wd}
  W_d=G_vV_d,  
\end{equation} where the coefficient $G_v$ is the energy required per unit volume, and can be considered as an effective material parameter. We emphasize that \eqref{Wd} accounts for the portion of work invested in altering the configuration of the matrix (i.e. its shape), in the reference frame.

We can now compare the total work invested in the growth process with or without damage. Before onset of damage at a given state, as prescribed by the set $(V_0, V(t),\bm{X}_{b},\bm{X}_{bI})$, the invested energy follows directly from \eqref{Lmin}, hence we write 
\begin{equation}\label{We}
W_{e}=\mathcal{L}_{min}(V_0, V(t),\bm{X}_{b},\bm{X}_{bI}).
\end{equation}
Upon damage, an instantaneous change in the configuration may occur, to arrive at a new effective void shape $\bm{X}_b^d$, without change in $\bm{X}_{bI}$. The total work invested in the deformation can  be decomposed into two parts; the work needed for shape change \eqref{Wd},   and for deformation from the new configuration, namely
\begin{equation}\label{Wed}
W_{ed}=W_d+\mathcal{L}_{min}(V_0, V(t),\bm{X}_{b}^d,\bm{X}_{bI}).
\end{equation}
For the material system to instantaneously transition to a damaged state with the same volume and stress-free shape of the inclusion $(\bm{X}_{bI},V(t))$, the mechanical process must obey the thermodynamic inequality 
\begin{equation}\label{ineq}
    \mathcal{D}=W_e-W_{ed}\geq0, 
\end{equation}
where $\mathcal{D}$ is the energy dissipated in the transition.  This implies that damage may initiate once $W_e=W_{ed}$.

To examine this limit, it is instructive to define the auxiliary function   
\begin{equation}\label{G_fun}
    \mathcal{G}(V_0, V(t),\bm{X}_{b},\bm{X}_{bI},\bm{X}_{b}^d)=\frac{\Delta\mathcal{L}_{min}}{V_d},
\end{equation}
where  \begin{equation}\label{DLmin}
  \Delta\mathcal{L}_{min}=\mathcal{L}_{min}(V_0, V(t),\bm{X}_{b},\bm{X}_{bI})-\mathcal{L}_{min}(V_0, V(t),\bm{X}_{b}^d,\bm{X}_{bI}),  
\end{equation}and the volume expansion is defined  as  $\Delta V=V(t)-V_0$.  

\begin{figure}[h]
  \makebox[\textwidth][c]{\includegraphics[width=1\textwidth]{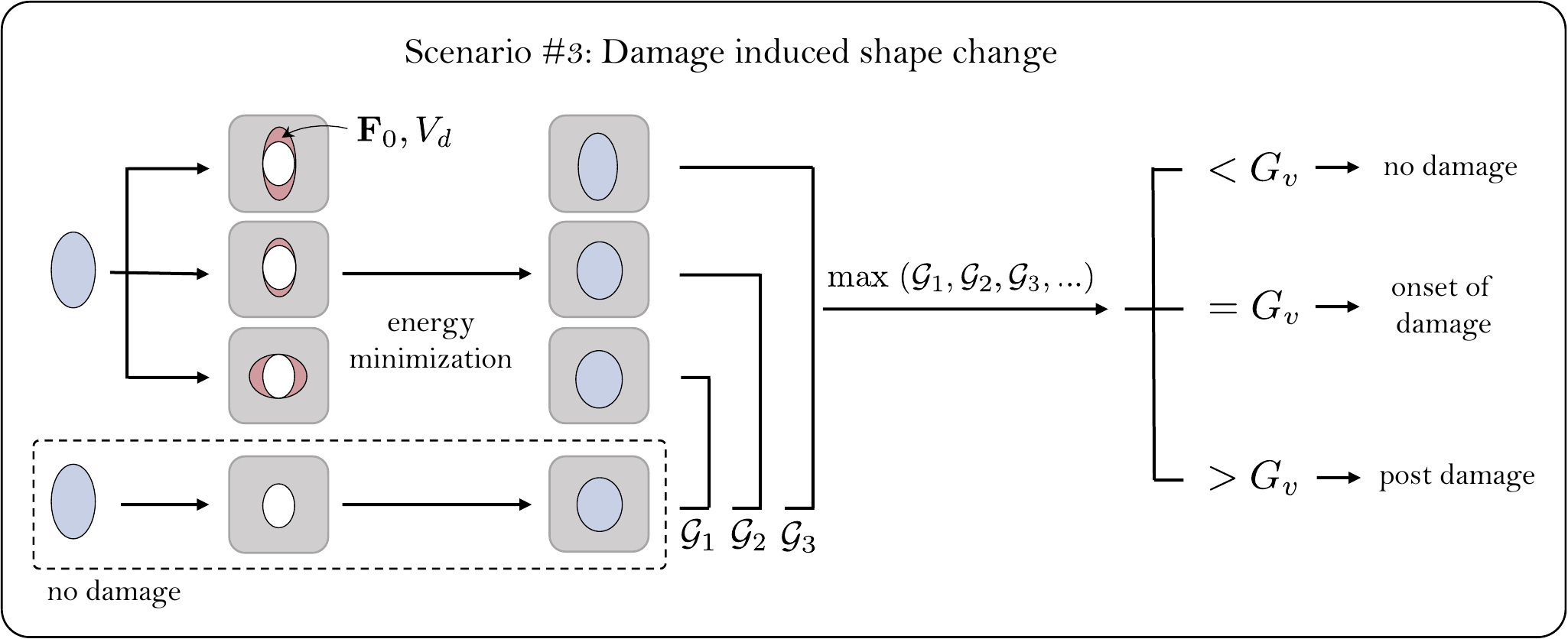}}%
  \caption{Schematic of Scenario \#3.}\label{fig:S3}
\end{figure}

If damage onset occurs at a critical expansion $\Delta V_{c}$ with a known shape of the inclusion $\bm{X}_{bI}$, then according to the inequality \eqref{ineq} we can write 
\begin{equation}\label{damage_cases}
    \begin{aligned}
        &\Delta V<\Delta V_c:  &\underset{\bm{X}_b^d}{\rm max}~{\mathcal{G}}<G_v \quad &\text{(no damage)}\\
         &\Delta V=\Delta V_c:  &\underset{\bm{X}_b^d}{\rm max}~{\mathcal{G}}=G_v \quad &\text{(onset of  damage)}\\
         &\Delta V>\Delta V_c:  &\underset{\bm{X}_b^d}{\rm max}~{\mathcal{G}}>G_v\quad  &\text{(post  damage)}.
    \end{aligned}
\end{equation}
This process of determining if the system is damaged or not is illustrated in Fig. \ref{fig:S3}. To understand how damage can affect changes in the observed shape, we will focus our attention to determining $\bm{X}_b^d$ at onset of damage, and the corresponding deformed shape $\bm{x}_b^d$. Before we can do so, it remains to prescribe a functional form of $V_d=V_d(\mathbf{F}_0)$. A simplistic  estimate of the damaged volume  can be chosen as the minimal region that encloses the initial void, but has the  shape of the effective damaged void, as illustrated schematically by the red shaded regions in Fig. \ref{fig:S3}. This assumption complies with the basic physical requirement that if there is no shape change there is no damaged volume (i.e. $V_d(\mathbf{I})=0$). 

In restricting our attention to elliptic stress-free shapes of the inclusion and the matrix, the set $(\bm{X}_b,\bm{X}_{bI}, \bm{X}_b^d)$ simplifies to the corresponding aspect ratios $(\Phi,\Phi_I, \Phi_d)$. The volume of the minimal concentric elliptic region that encloses an ellipse with the initial aspect ratio $\Phi$, and has the damaged aspect ratio $\Phi_d$ is readily obtained as $V_0\Omega(\Phi,\Phi_d)$, where  \begin{equation}\label{Omega}
 \Omega(\Phi,\Phi_d)=  \begin{cases}
        \Phi_d/\Phi-1 \quad \Phi_d \geq\Phi\\
        \Phi/\Phi_d-1 \quad \Phi_d <\Phi.
    \end{cases}
\end{equation} 

Since damage can be directional and thus partial (i.e. $G_v$ is the upper bound on the energy per unit volume), a similar effective shape can be achieved with different regions of partial damage, and will be chosen based on the particular constitutive model. To account for this effect we introduce an additional effective material parameter, $n$, and  define the damaged volume as  \begin{equation}\label{Vd}
    V_d=V_0\Omega^n.
\end{equation}  Thence, using \eqref{Lmin_ell}, we can rewrite \eqref{G_fun} as
\begin{equation}\label{G_fun_ellipse}
    \mathcal{G}(\Phi,\Phi_d)=\frac{\overline{\Delta\mathcal{L}}_{min}(\Phi_d-\Phi)}{V_0\Omega(\Phi,\Phi_d)^n},
\end{equation} where without loss of generality we replace ${\Delta\mathcal{L}}_{min}(\Phi,\Phi_d)=\overline{\Delta\mathcal{L}}_{min}(\Phi-\Phi_d)$ and use the shorthand notation to write the above relation for a given set $(V_0,V(t),\Phi_I)$.

\bigskip

Now, we use \eqref{G_fun_ellipse}  to consider the physical bounds on the parameter $n$: 

\noindent First, if \underline{ $\bm{n>1}$,} through Taylor expansion of $\overline{\Delta\mathcal{L}}_{min}$,  it can be shown that  ${\rm max}~\mathcal{G}\to\infty$ at $\Phi_d=\Phi$ for all $(V_0,V(t),\Phi_I)$. Hence, damage will always initiate immediately at onset of loading and will not induce shape change. This result is considered to be nonphysical.

Next, if $\overline{\Delta\mathcal{L}}_{min}$ is concave (i.e. $\overline{\Delta\mathcal{L}}''_{min}<0$), we can write the inequality\footnote{Here we use the inequality $\overline{\Delta\mathcal{L}}_{min}(\Phi-\Phi_d)\leq\overline{\Delta\mathcal{L}}_{min}(0)+\overline{\Delta\mathcal{L}}'_{min}(0)\cdot(\Phi-\Phi_d)$.} 
\begin{equation}
    \mathcal{G}(\Phi,\Phi_d)\leq  \frac{\overline{\Delta\mathcal{L}}'_{min}(0)}{V_0} \begin{cases}
        ~~\Phi^n(\Phi_d-\Phi)^{1-n}& \quad \Phi_d \geq\Phi\\
        -\Phi_d^n(\Phi-\Phi_d)^{1-n}&\quad \Phi_d <\Phi.
    \end{cases}
\end{equation}

\noindent Accordingly, if \underline{$\bm{n=1}$,} then the maximum,  ${\rm max}~{\mathcal G}=\Phi^n\overline{\Delta\mathcal{L}}'_{min}(0)/V_0$,  appears at $\Phi=\Phi_d$ and onset of damage will not lead to changes in shape. 

\noindent If \underline{$\bm{n<1}$,} a solution for which ${\rm max}~{\mathcal G}=G_v$ and  $\Phi_d\neq\Phi$ may exist. Thence, using \eqref{G_fun_ellipse}, at onset of damage  we can find the corresponding effective shape from $\Phi_d={\rm arg}~\underset{\bar{\Phi}_d}{\rm max}~\mathcal{G}(\Phi,\bar{\Phi}_d)$ and the observed shape, $\phi$, is obtained from \eqref{Lmin_ell},  by setting $\Phi=\Phi_d$.

Overall, our minimal damage model introduces two material parameters whose physical values are in the ranges $G_v\geq0$ and $n\leq1$. The sensitivity of the response to these parameters will become apparent in view of results in Section \ref{sec:results}.

\bigskip

\noindent \textbf{  \textit{Scenario \#4: Matrix remodeling in response to growth}}

 An alternative adaptation that may manifest in the confining medium, in response to the expanding inclusion, is remodeling. As in Scenario \#2, we can  postulate  that morphogenesis of the matrix  will  proceed  in  a  way  to  minimize the total mechanical energy of the system. Hence, if the matrix were able to adapt the volume of the void (in its stress-free state) to accommodate the growing body, the system would remain stress-free. On the other hand,  if changes in volume are prohibited, volume preserving changes in configuration, can  potentially allow the two body system to further alleviate the strains of growth, or in other words, the amount of elastic energy in the system. 

\begin{figure}[h]
  \makebox[\textwidth][c]{\includegraphics[width=1\textwidth]{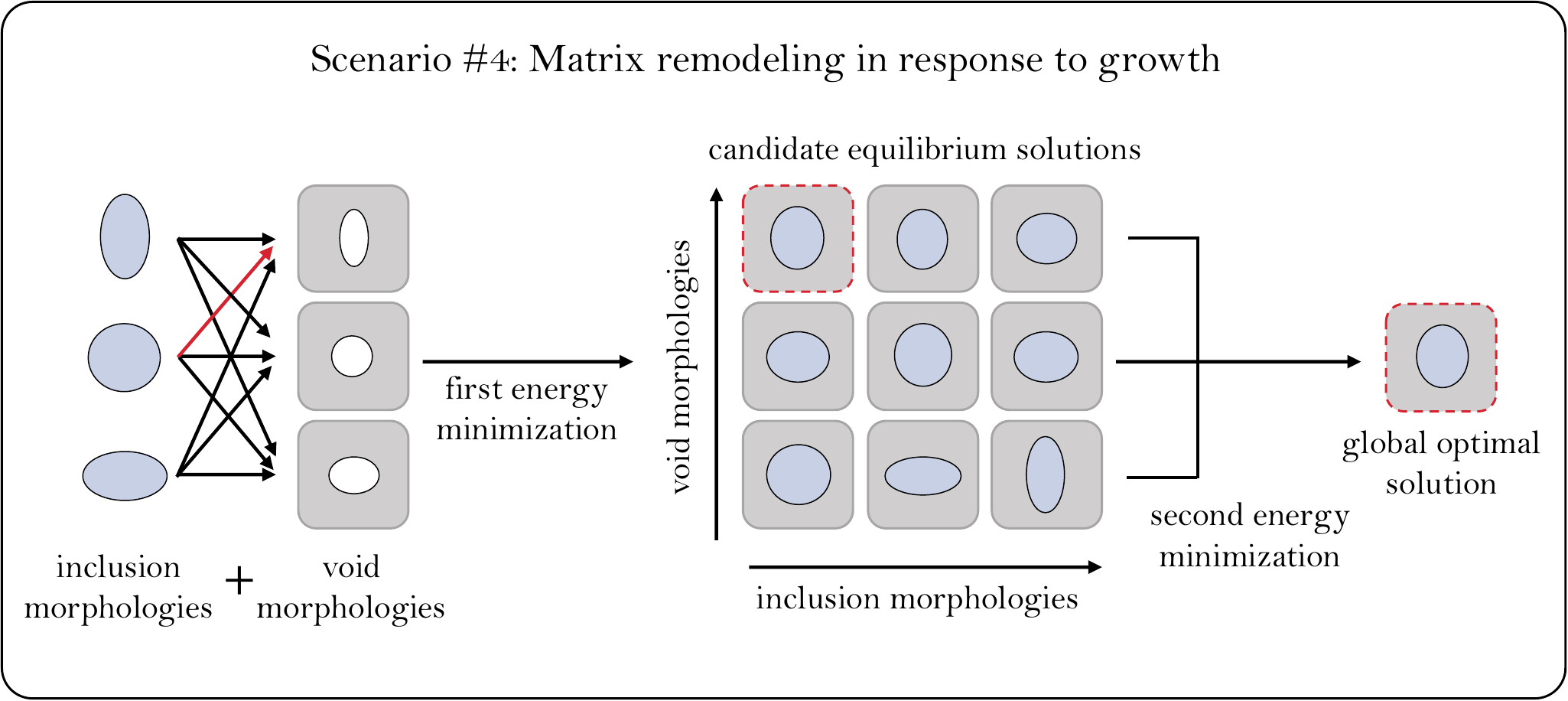}}%
  \caption{Schematic of Scenario \#4.}\label{fig:S4}
\end{figure}

To capture this growth Scenario  we extend Scenario \#2, to include the stress-free shape of the void ${\bm{{X}}_{b}}$ as an additional  parameter.  Accordingly, \eqref{axiomA} takes the form 
\begin{equation} \label{axiomA_S4}
    (\bm{X}_{b},\bm{X}_{bI})= {\rm arg} ~  \underset{(\bm{\bar X}_{b},\bm{\bar X}_{bI})}{\rm min} ~\left\{{\rm min} ~    \mathcal{L}(\mathbf{\chi},\mathbf{\chi}_I,{\mathbf F},{\mathbf F}_I;\bm{x}_b({\bm{\bar{X}}_{b}},{\bm{\bar{X}}_{bI}}))\right\},
\end{equation}
 and specializing this solution  procedure to the case of elliptic stress-free configurations, reads
\begin{equation} \label{axiomA_S4el}
    (\Phi,\Phi_I)= {\rm arg} ~  \underset{(\bar{\Phi},\bar{\Phi}_I)}{\rm min} ~\left\{{\rm min} ~    \mathcal{L}(\mathbf{\chi},\mathbf{\chi}_I,{\mathbf F},{\mathbf F}_I;{ \phi}(\bar{\Phi},\bar{\Phi}_I))\right\}.
\end{equation}
Even for the simplified geometry, the above solution procedure requires a double optimization scheme on a two-dimensional space, as illustrated in Fig. \ref{fig:S4}.

Solutions procedures for this growth scenario, as well as Scenarios \#2 and \#3, rely on  equilibrium solutions for arbitrary aspect ratios and volume expansions, obtained in Scenario \#1. In this work, we will use two different approximate methods to obtain solutions for Scenario \#1, as detailed in the next section.





\section{Equilibrium solutions for prescribed anisotropic growth}\label{sec:equil}
The  optimization schemes theorized for the different growth scenarios in the previous section, rely on the knowledge of equilibrium solutions for arbitrary shapes and volume expansions (i.e. Scenario \#1). Then optimization is conducted among all possible configurations at finite levels of volume expansion. While solution for a single prescribed anisotropic growth process within a finite deformation continuum framework can be challenging, optimization among all possible configurations can become intractable.  

In  this section, we present two alternative modeling approaches in which we restrict our attention to elliptic stress-free shapes of both the inclusion and the void.  Each of these methods has its own limitations: First we will employ a minimal analytical approach that applies stringent kinematic assumptions to reduce the number of unknown field variables, thus  permitting quasi-analytical treatment, but reducing the accuracy of our results green \citep{ericksen1954deformations,ericksen1955deformations}. This solution applies for situations in which sliding is permitted at the interface (i.e there is no transmission of shear traction). Then, we employ finite element method, that allows for high accuracy, but requires long computation times. This solution considers a strictly bonded interface. In the following section (Section \ref{sec:results}) we will compare the results obtained using the different methods and will consider also their agreement with exact solutions obtained at the linear limit.

\subsection{Quasi-Analytical Method  (QAM)}

\noindent to describe this growth process, we begin by making a simplifying assumption on the deformation field. 
We consider situations in which the different shapes of the void and the inclusion can be represented reasonably well as ellipses in both their stress-free and deformed configurations. 

\bigskip 

\noindent \textbf{Reference configuration.} In the undeformed state, we define the cartesian coordinate system $(X,Y,Z)$ whose origin is at the center of the void. The former two coordinates are along the minor and major axis of the ellipse, respectively, and the latter points out of the plane to  complete\ the right-handed orthogonal set (Fig. \ref{illus}). Now, considering the undeformed states of both the matrix and the inclusion, we can parameterize the field as a continuum of concentric ellipses, which have the same aspect ratio as the undeformed void and inclusion, respectively. 
In the plane, this parameterization is chosen such that an   ellipse is identified by the location of its intersection  with the minor axis,  $\Lambda$, and particles along its  circumference are identified by their location $\Theta$. Mathematically, this implies \begin{equation}
{\bm{X}}(\Lambda,\Theta,Z;\Phi):=(\Lambda \cos\Theta, \Phi \Lambda\sin\Theta,Z), 
\end{equation} and thus, in the reference state, the matrix occupies the region\begin{equation}
A\leq \Lambda<\infty,\qquad 0\leq\Theta<2\pi,\qquad-\infty<Z<\infty,
\end{equation}whereas the inclusion occupies the region\begin{equation}
0\leq \Lambda<A_{I},\qquad 0\leq\Theta<2\pi,\qquad-\infty<Z<\infty.
\end{equation}

\noindent \textbf{Current configuration.} Upon deformation we assume that these ellipses are mapped into new ellipses.     Hence, given the deformed cartesian  coordinate system  $(x,y,z)$, we write the parameterization in terms of the deformed aspect ratio $\varphi$, as
\begin{equation}\label{x}
\bm{x}(\alpha,\theta,Z;\varphi):=(\alpha \cos\theta, \varphi \alpha\sin\theta,z), 
\end{equation}and thus, in the reference state, the matrix occupies the region\begin{equation}
a\leq \alpha<\infty,\qquad 0\leq\theta<2\pi,\qquad-\infty<z<\infty,
\end{equation}whereas the inclusion occupies the region\begin{equation}
0\leq \alpha<a,\qquad 0\leq\theta<2\pi,\qquad-\infty<z<\infty.
\end{equation}

\noindent \textbf{Placement map and its gradient.} Next, we define a placement map that assigns an undeformed ellipse to its new deformed position, namely $\bm{x}={\bm{\chi}}({\bm{X}},t)$. Notice that this mapping varies with time;  the consequence of this evolution will become apparent soon. Within the confines of our assumption that ellipses remain ellipses in a plane-strain setting, the most general transformation that excludes any rigid body translations or rotations and permits analytical treatment\footnote{Note that, conceptually,  a more general transformation can include changes in $\theta$ in the form $\theta=\Theta+\vartheta(\Lambda,\Theta)$, where admissibility implies the additional requirement of periodicity, such that  $\vartheta(\Lambda,n\pi/2)=0$, for $n=0,1,2$. However, such an assumption renders an analytical derivation untractable.  }, is 
\begin{equation}
\alpha=\alpha(\Lambda), \qquad \varphi=\varphi(\Lambda),\qquad \theta=\Theta,\qquad z=Z,
\end{equation}

Provided this transformation we can now write the mapping as \begin{equation}
{\bm{\chi}}({\bm{X}},t)=\bm{x}(\alpha(\Lambda),\varphi(\Lambda),\Theta,Z),
\end{equation} and its gradient is derived by invoking the chain rule to write
\begin{equation}\label{F0}\begin{split}
\mathbf{F}=\frac{\partial \bm x}{\partial \bm X}=&\frac{\partial \bm x}{\partial \alpha}\otimes  \left(\frac{\partial \alpha}{\partial \Lambda} \frac{\partial \Lambda}{\partial \bm X}\right)+\frac{\partial \bm x}{\partial \varphi}\otimes \left(\frac{\partial \varphi}{\partial \Lambda} \frac{\partial \Lambda}{\partial \bm X}\right) \\& +\frac{\partial \bm x}{\partial \theta}\otimes \left(\frac{\partial \theta}{\partial \Theta} \frac{\partial \Theta}{\partial \bm X}\right) + \frac{\partial \bm x}{\partial z}\otimes\left(\frac{\partial z}{\partial Z} \frac{\partial Z}{\partial \bm X}\right).
\end{split}\end{equation}
To specify this deformation gradient in Cartesian coordinates, we first write the covariant vectors of the curvilinear coordinates in the current frame, as     \begin{equation}\label{covar}
\frac{\partial \bm x}{\partial \alpha}=\left(\cos\theta,\varphi\sin\theta,0\right),\quad \frac{\partial \bm x}{\partial \theta}=\left (-\alpha\sin\theta ,\varphi \alpha\cos\theta,0\right), \quad \frac{\partial \bm x}{\partial z}=\left(0,0,1\right), \quad \frac{\partial \bm x}{\partial \varphi}=\left(0,\alpha\sin\theta,0\right),
\end{equation} 
and the contravariant vectors, in the reference frame, as     
\begin{equation}\label{contra} \frac{\partial \Lambda}{\partial \bm X}=\frac{1}{\Phi}\left(\Phi\cos\Theta,\sin\Theta,0\right),\quad \frac{\partial \Theta}{\partial \bm X}=\frac{1}{\Phi \Lambda}\left(-\Phi\sin\Theta,\cos\Theta,0\right), \quad 
\frac{\partial Z}{\partial \bm X}=\left(0,0,1\right.).   
\end{equation}Here the vectors are written with their components in the directions of the cartesian unit vectors $( \mathbf{e}_{\rm x},\mathbf{e}_{\rm y},\mathbf{e}_{\rm z})$, which are identical in the reference and current frames. Now, by inserting \eqref{covar} and \eqref{contra} into \eqref{F0} we rewrite the deformation gradient as\begin{equation}\label{F}\begin{split}
\mathbf{F}&= \left(\bar\lambda\sin^{2}\Theta+\lambda\cos^{2}\Theta \right)\mathbf{e}_{\rm x}\otimes \mathbf{e}_{\rm x}+\frac{1 }{\Phi}(\lambda-\bar\lambda)\sin\Theta\cos\Theta~ \mathbf{e}_{\rm x}\otimes \mathbf{e}_{\rm y}\\&+\varphi(\lambda-\bar\lambda+\beta)\sin\Theta\cos\Theta~ \mathbf{e}_{\rm y}\otimes \mathbf{e}_{\rm x}+\frac{\varphi }{\Phi}\left((\lambda+\beta)\sin^{2}\Theta+\bar\lambda\cos^{2}\Theta\right)\mathbf{e}_{\rm y}\otimes \mathbf{e}_{\rm y}+\mathbf{e}_{\rm z}\otimes \mathbf{e}_{\rm z},
\end{split}\end{equation} 
where  we have substituted the shorthand notations \begin{equation}\label{defs}
\lambda=\frac{{\rm d} \alpha}{{\rm d} \Lambda},\qquad \bar\lambda=\frac{\alpha}{\Lambda},\qquad \beta=\frac{\alpha}{\varphi}\left(\frac{{\rm d} \varphi}{{\rm d} \Lambda}\right).
\end{equation}

For future use,  the first invariant of the left Cauchy Green deformation tensor $\mathbf{B}=\mathbf{F}\mathbf{F}^{\rm T}$ is readily written, after some algebra,  as
\begin{equation}\label{I1}\begin{split}
{ I_1}=&\left(\frac{\varphi^{2} }{\Phi^{2}}(\lambda+\beta)^{2}+\bar\lambda^{2}\right)\sin^{4}\Theta+\left(\lambda^{2}+\frac{\varphi^{2} }{\Phi^{2}}\bar\lambda^{2}\right)\cos^{4}\Theta\\&+\left(\frac{1 }{\Phi^{2}}(\lambda-\bar\lambda)^{2}+\varphi^{2}(\lambda-\bar\lambda+\beta)^{2}+2\lambda\bar\lambda+2\frac{\varphi^{2} }{\Phi^{2}}(\lambda+\beta)\bar\lambda\right)\sin^{2}\Theta\cos^{2}\Theta +1,
\end{split}\end{equation}and the third invariant $J=\det\mathbf{F}$, simplifies to \begin{equation}\begin{split}
J&=\frac{\varphi }{\Phi } \bar\lambda(\lambda+\beta\sin^{2}\Theta).  
\end{split}\end{equation}

\noindent \textbf{Boundary and compatibility conditions.} While the above mapping   applies for both the matrix and the inclusion, the field must be solved separately for each, to obtain $\alpha(\Lambda)$ and $\alpha_I(\Lambda)$, respectively. Compatibility of the two bodies, in the current frame, follows from \eqref{compt}, which for the present setting implies  the boundary condition\begin{equation}\label{bc1}
a=\alpha(A)=\alpha_{I}(A_{I}).
\end{equation} Similarly, the deformed aspect ratio $\varphi(\Lambda)$ may vary in  the matrix. In the inclusion, we anticipate a uniform deformation, such that $\varphi_I(\Lambda)\equiv\phi$. Accordingly, we write an additional compatibility condition \begin{equation} \label{bc2}
\phi=\varphi_I=\varphi(A). 
\end{equation}The remaining boundary, is the remote boundary of the matrix. It is expected that  the remote field is undisturbed by the presence of the growing inclusion\footnote{In this work we limit our discussion to situations in which the  inclusion remains small compared to the size of the medium within it is growing. It is straightforward to modify this boundary condition to account for  finite matrix  dimensions, however within the assumptions of the present framework, this matrix must have an elliptic shape.}  and is thus uniform, as is the deformation field in the inclusion. Hence, for the present setting the remote boundary condition \eqref{RC0} specializes to \begin{equation}\label{bc3}
\lambda(\Lambda\to\infty)\to\lambda_\infty,
\end{equation}if the remote stretch ($\lambda_\infty$) is imposed. Or if the remote field is undisturbed, we have $\lambda_{\infty}=1$. The corresponding aspect ratio in the remote field then follows from the requirement of incompressibility.      

\smallskip

\noindent \textbf{Incompressibility constraint.} Limiting our formulation to incompressible materials, we require that concentric elliptical regions within the material preserve their volume.  For the inclusion, this implies \begin{equation}\label{incomp_incl}
\pi\phi \alpha^2=\pi\Phi_{I} \Lambda^2\quad\to \quad \bar\lambda=\sqrt{\Phi_I/\phi}.
\end{equation} 
whereas the total volume of the inclusion is $V(t)=\pi \phi a^2=\pi\Phi_{I} A_{I}^2$ .
In the matrix, the additional volume due to expansion of the inclusion must be accounted for to write \begin{equation}\label{incomp_mat}
\pi\varphi \alpha^2-\pi\Phi \Lambda^2=\Delta V\quad\to\quad \varphi=\frac{\Delta V+\pi\Phi \Lambda^2}{\pi \alpha^2}.
\end{equation}where the total volume change is $ \Delta V=V(t)-V_0=\pi(\Phi_{I} A^{2}_{I}-\Phi A^{2})$.

Note that both \eqref{incomp_incl} and \eqref{incomp_mat}  identically satisfy the integral requirement of incompressibility \begin{equation}
\int\displaylimits_{\Lambda_1}^{\Lambda_2}\int\displaylimits_0^{2\pi}(J-1)\Lambda{\rm d}\Theta {\rm d}\Lambda=\frac{2\pi }{\Phi}\int\displaylimits_{\Lambda_1}^{\Lambda_2}\left[\varphi \bar\lambda\left(\lambda+\frac{\beta}{2}\right)-1\right]\Lambda{\rm d}\Lambda=0, 
\end{equation}which implies \begin{equation}\label{int_detF}
\varphi \bar\lambda\left(\lambda+\frac{\beta}{2}\right)=1\quad\to\quad\beta=2\left(\frac{1}{{\varphi \bar\lambda}}-\lambda\right),
\end{equation}in any elliptical subregion $\Lambda\in[\Lambda_1,\Lambda_2]$. 

By inserting the above relations in \eqref{F} and \eqref{I1} and recalling the definitions \eqref{defs},  the deformation gradient and the first invariant in the matrix can be written as functions of\footnote{Here we have introduced the shorthand notation $\alpha'={\rm d} \alpha/{\rm d} \Lambda$.} $(\Lambda,\alpha,\alpha',\Theta)$, for a given set of model parameters $(\Phi,A,V_0,V(t))$, namely
\begin{equation}\label{FI1_mat}
\mathbf{F}=\mathbf{F}(\Lambda,\alpha,\alpha',\Theta;\Phi,A,V_0,V(t)), \quad I_1=I_1(\Lambda,\alpha,\alpha',\Theta;\Phi,A,V_0,V(t)).
\end{equation}For the inclusion, combining \eqref{bc1},\eqref{bc2}, and \eqref{incomp_incl}  implies $\lambda=\bar\lambda=a/A_{I}$ and $\varphi=\phi=\Phi_I(A_{I}/a)^{2}$, such that $\beta=0$. Hence, the deformation gradient and the first invariant reduce to the expected results \begin{equation}\label{FI1_incl}\begin{split}
\mathbf{F}&=\frac{a}{A_{I}}\mathbf{e}_{\rm x}\otimes \mathbf{e}_{\rm x}+\frac{A_{I} }{ a}\mathbf{e}_{\rm y}\otimes \mathbf{e}_{\rm y}+\mathbf{e}_{\rm z}\otimes \mathbf{e}_{\rm z},\quad I_1=\left(\frac{a}{A_{I}}\right)^2+\left(\frac{A_{I}}{a}\right)^2+1,
\end{split}\end{equation}which, given the undeformed semi-minor axis - $A_I$,  depends only on  the deformed value -  $a$. 
 
\smallskip

\noindent \textbf{Summary of kinematic assumptions.}  Before proceeding  to consider the kinetics of this problem, we summarize the two  assumptions that have been used to define the kinematics:
\begin{enumerate}
\item We have restricted the deformation such that  concentric ellipses in the reference frame are mapped into new concentric ellipses in the current frame. This assumption entails the limitation that the shapes of the void and the inclusion in the reference frame can be represented reasonably well as ellipses. 

\item  We limit our discussion to incompressible materials. This volume constraint is imposed in an averaged sense, such that any elliptic sub-region within the matrix or the inclusion preserves its volume. \end{enumerate} 
At the limit of axially symmetric expansion, these assumptions are identically satisfied and the results of our model are exact. For non-circular expansion, the consequences of these assumptions, will be further examined in  comparison with numerical solutions and linear elastic results. 

\smallskip
\noindent\textbf{Equilibrium.}
With the kinematics fully defined, we now proceed to consider the potential energy of the two body system to examine possible  solutions, which minimize this energy. Recalling the total energy in \eqref{S0}, and in view of \eqref{FI1_mat},\eqref{FI1_incl}, we define the two auxiliary functions for mathematical convenience
\begin{equation}\label{psi_defs}
\psi(\Lambda,\alpha,\alpha';\Phi,A,V_0,V(t))=\frac{\Lambda}{2\pi}\int\displaylimits_{0}^{2\pi}\Psi(\mathbf F) {\rm d}\Theta, \qquad \psi_I(a,A_{I})=\frac{1}{2\pi}\int\displaylimits_{0}^{2\pi}\Psi_{I}(\mathbf F) {\rm d}\Theta.\end{equation}These definitions are now inserted into \eqref{S0} to rewrite the energy as a functional, which depends  on one unknown function, $\alpha(\Lambda)$, for a given set $(\Phi,A,V_0,V(t))$,  in the form\begin{equation}\label{E1}
\frac{\mathcal{L}(\alpha(\Lambda))}{2\pi\Phi\mu_M}=\int\displaylimits_{A}^{\infty}\psi(\Lambda,\alpha,\alpha';\Phi,A,V_0,V(t)){\rm d}\Lambda +\frac{\mu_I}{\mu_M}\left(\frac{V(t)}{2\pi\Phi}\right)\psi_{I}(a,A_{I}).
\end{equation}
Note that in this formulation, provided the kinematic assumption that ellipses remain ellipses, the constraint in \eqref{S0} is identically satisfied, and sliding between the two surface is permitted. 

Now, following the basic principles of calculus of variations\footnote{For completeness, we show here the full derivation.}, we seek a function, $\alpha(\Lambda)$, which  makes  $\mathcal{L}$ stationary, and obeys both the compatibility requirements and boundary conditions. To this end, we consider variations of  $\alpha$ in the form $\hat \alpha(\Lambda)=\alpha(\Lambda)+\varepsilon\eta(\Lambda)$  where $\varepsilon$ is a small constant and  $\eta(\Lambda)$   is an arbitrary function, which must vanish in the remote field, i.e.   $\eta(\Lambda\to\infty)=0$, to accommodate the displacement boundary condition \eqref{bc3}, which implies a prescribed value of $\alpha$. We insert this variation in \eqref{E1} and require an extremum such that \begin{equation}
\delta \mathcal{L}=\underset{\varepsilon\to 0}{\lim}\frac{{\rm d}\mathcal{L}(\alpha+\varepsilon \eta)}{{\rm d}\varepsilon}=0.
\end{equation}Explicitly, this reads
\begin{equation}\begin{split}
\frac{\delta \mathcal{L}}{2\pi\Phi\mu_M}&=\int\displaylimits_{A}^{\infty}\left(\frac{\partial\psi }{\partial \alpha}\eta+\frac{\partial\psi }{\partial \alpha'}\frac{{\rm d}\eta}{{\rm d}\Lambda}\right){\rm d}\Lambda + \frac{\mu_I}{\mu_M}\left(\frac{V(t)}{2\pi\Phi}\right)\frac{\partial\psi_{I}}{\partial a}\eta(A)\\
&=\int\displaylimits_{A}^{\infty}\left(\frac{\partial\psi }{\partial \alpha}-\frac{{\rm d}}{{\rm d}\Lambda}\left(\frac{\partial\psi }{\partial \alpha'}\right)\right){\rm \eta d}\Lambda -\left(\frac{\partial\psi }{\partial \alpha'}\bigg|_{\Lambda=A}-\frac{\mu_I}{\mu_M}\left(\frac{V(t)}{2\pi\Phi}\right)\frac{\partial\psi_{I}}{\partial a}\right)\eta(A)=0,
\end{split}
\end{equation}
where we have already accounted for the vanishing value of $\eta$ at the remote boundary. 

Finally, for this expression to vanish identically for any arbitrary variation, we require that both terms vanish separately, to write\begin{equation}\label{interface_cond}
\frac{\partial\psi }{\partial \alpha}-\frac{\rm d}{{\rm d}\Lambda}\left(\frac{\partial\psi }{\partial \alpha'}\right)=0,\quad \text{and}\quad \frac{\partial\psi }{\partial \alpha'}\bigg|_{\Lambda=A}=\frac{\mu_I}{\mu_M}\left(\frac{V(t)}{2\pi\Phi}\right)\frac{\partial\psi_{I}}{\partial a}.
\end{equation}
The first relation is a second order nonlinear differential equation in $\alpha(\Lambda)$, which  serves as the governing equation in the matrix. The latter is a natural boundary condition that applies at the interface between the matrix and the inclusion, and replaces the minimization in \eqref{axiomB_el}. From this boundary condition it is seen that the present reduced model affords only one \textit{generalized reaction force} acting between the two bodies. This reaction force is directly proportional to the deformation of the inclusion (i.e. $\partial \psi_I/\partial a$).  

Provided a constitutive response function, the above second order nonlinear differential equation is integrated numerically and the two boundary conditions, \eqref{bc3}, \eqref{interface_cond}$^2$, are enforced through a shooting method, along with the compatibility condition \eqref{bc1}. Upon obtaining an equilibrium solution, the energy in the system  can be calculated by inserting the  equilibrium field in \eqref{E1} and performing integration.

\smallskip

\noindent \textbf{Specific constitutive model.} To examine solutions, in this work we will use the neo-Hookean material model for both the matrix and the inclusion. Hence, we have  $\Psi=\frac{1}{2}\left( I_{1}-3\right)$. By inserting the  first invariants \eqref{FI1_mat},\eqref{FI1_incl} in \eqref{psi_defs} and integrating over the angular coordinate, we obtain \begin{equation}\begin{split}
\psi=\frac{\Lambda}{16}\bigg(&3(\varphi^{2}(\lambda+\beta)^{2}+\Phi^{2}\bar\lambda^{2}+\Phi^{2}\lambda^{2}+\varphi^{2}\bar\lambda^{2})+(\lambda-\bar\lambda)^{2}
\\&+\Phi^{2}\varphi^{2}(\lambda-\bar\lambda+\beta)^{2}+2\Phi^{2}\lambda\bar\lambda+2\varphi^{2}{ }{}(\lambda+\beta)\bar\lambda-16\bigg),
\end{split}\end{equation}for the matrix, and \begin{equation}\label{psiI}
\psi_{I}=\frac{1}{2}\left(\frac{a^{2}-A_{I}^2}{a A_{I}}\right)^2,
\end{equation} for the inclusion. 

\smallskip
\noindent \textbf{Analytical result for Scenario \#2.} Recall that in Scenario \#2 we seek the shape of the inclusion that minimizes the energy in the system. Among the set $(V(t), \Phi_I, A_I)$ two independent parameters are sufficient to describe the shape and size of the inclusion, provided the relation $V(t)=\pi\Phi_IA_I^2$. Hence, for a given volume, $V(t)$, with finite stiffness ratio $\mu_I/\mu_M$ the minimization in \eqref{axiomA_el} translates to  \begin{equation}
    \frac{\partial \mathcal{L}_{min}}{\partial A_I}=0 \quad \xrightarrow[]{\eqref{E1}} \quad \frac{\partial \psi_I}{\partial A_I}=0\quad \xrightarrow[]{\eqref{psiI}} \quad A_I=a.
\end{equation} 
Here we obtain that for this reduced model, by allowing the  inclusion to select a shape that minimizes the energy of the entire system, it selects a response of a perfectly rigid inclusion (with $\psi_I=0$), independent of the stiffness ratio. For this simple model, at this limit, the response of a rigid inclusion, and that of  fluid expansion in Scenario \#1  are indistinguishable and imply vanishing of the generalized reaction force \eqref{interface_cond}$^2$.   We will further discuss the limitations of this result in the next section.


\subsection{Finite Element Model (FEM)}
While the quasi-analytical model detailed above can provide insights into the mechanical problem at hand with minimal computational effort, the kinematic assumptions that it employs inevitably limit its quantitative accuracy. An alternative and more accurate solution  can be obtained via the direct application of a finite element method, as detailed in this section. However, this approach also has some limitations. First, the computation is vastly more time consuming. Recall that the different scenarios that we consider require selection among various candidate solutions  (Section \ref{sec:def}). Using the FEM for just one expansion into the nonlinear range (i.e. Scenario \#1) is already  computationally-intensive, let alone the computation of a continuum of expansion scenarios among which one optimal solution is selected. Second, relative motion between points at the interface even further complicates the solution procedure. For this reason, in the FEM solution, we will simplify  the interface condition to be strictly bonded. Additionally, to reduce computation times, we will examine solutions at discrete values of volume expansion. 

All of our finite element solutions initiate from a stress-free compatible state\footnote{Note that in principle there need not be such a state. }, such that $V(0)=V_0$ with $\Phi_I(0)=\Phi$ and $A_I(0)=A$. It is this initial state that defines the bonding between material points at the interface. Then, expansion of the inclusion is prescribed at every time via the configurational tensor, $\mathbf{F}_0(t)$, which for elliptical stress-free shapes reads
\begin{equation}\label{FI1_incl_FEM}\begin{split}
\mathbf{F}_0=\frac{A_I}{A}\mathbf{e}_{\rm x}\otimes \mathbf{e}_{\rm x}+\frac{\Phi_IA_{I} }{\Phi A }\mathbf{e}_{\rm y}\otimes \mathbf{e}_{\rm y}+\mathbf{e}_{\rm z}\otimes \mathbf{e}_{\rm z}, \quad\text{where}\quad{\rm det}\mathbf{F}_0=\frac{V(t)}{V_0}.
\end{split}\end{equation}
This procedure is implemented in COMSOL 5.3a and employs a nearly incompressible neo-Hookean constitutive relation\footnote{The bulk modulus is set to $10^3$ times the shear modulus, and it is confirmed that volume changes are negligible.}.   
To simulate the response of an unbounded confining medium, we set the radius of the matrix to $10^3$ times the semi-major axis of the elliptical inclusion. For computational efficiency, only one quarter of the region is modeled and symmetry boundary conditions are imposed. A highly refined mesh is generated in the vicinity of the inclusion and it becomes coarser as the remote field is approached.  Mesh sensitivity analysis confirms that simulation results are mesh-size independent. 

In the next section, we compare results of the different modeling approaches to examine the morphogenesis of an inclusion as it grows in confinement following different growth Scenarios.

\bigskip
\section{Results and Discussion} \label{sec:results}

In this section, we will analyze results obtained for the various growth scenarios to answer Questions A and B posed in Section \ref{sec:def}. We use both the quasi-analytical and the finite element methods while limiting  our attention to elliptical stress-free shapes, as described above.

\begin{figure}[ht!]
  \centering
    \includegraphics[width=0.65\textwidth]{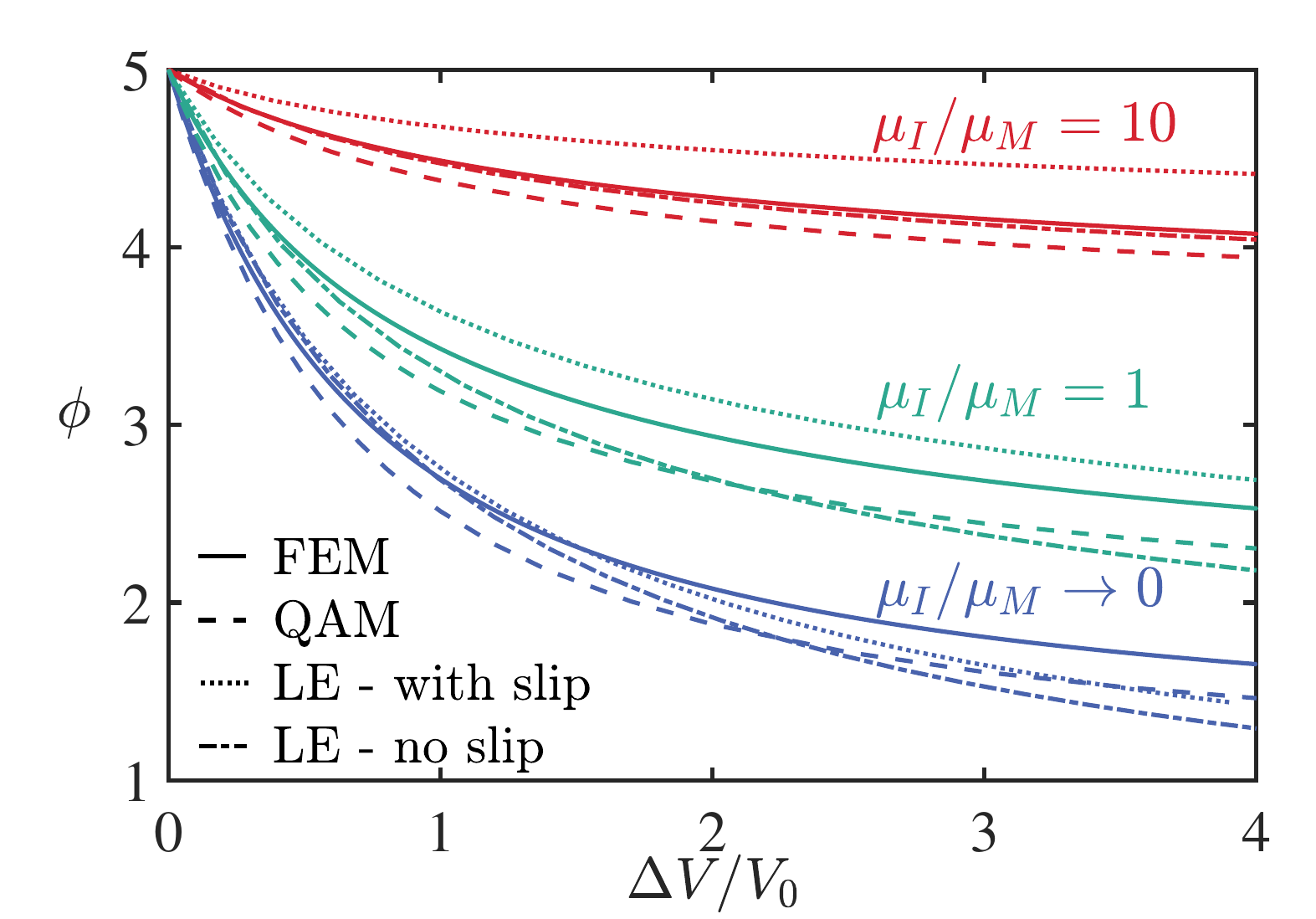}
  \caption{Scenario \#1 results for isotropic expansion with  ($\Phi=\Phi_I = 5$). The deformed shape of the inclusion is shown as a function of the expansion ratio using for different models that account for the two limits of the interface condition.}
  \label{scenario1_fig}
\end{figure} 

\bigskip
\noindent \textbf{Scenario \#1}. As a first example we consider a representative case where the void and the inclusion have the same stress-free aspect ratio $\Phi =\Phi_I =5$. 
The inclusion undergoes  isotropic expansion such that its volume  $V(t)$ is prescribed as it maintains its stress-free shape  $\Phi_I(t) =5$. 
This defines the set $(V_0, V,\Phi,\Phi_I)$ which completes the energy functional in \eqref{S00}.
Using energy minimization \eqref{axiomB_el}, we solve for the deformed shape, characterized by $\phi$. 
Results  in Fig. \ref{scenario1_fig} are shown for a broad range of stiffness ratios. 
 Recall that our QAM permits slipping at the interface while the  FEM approach has no slip, such that the strict bonding between the two bodies is defined by the initial state. Hence, for the purposes of comparison, we also include curves obtained using Linear Elastic inclusion theory (LE) for both the `no slip' and the `with slip' interface conditions. The `no slip` solution procedure follows the methodology outlined by  \cite{Eshelby1957TheDO}, based on the \textit{Eshelby tensor}, and specializes it for a two-dimensional elliptic cylindrical inclusion. The `with slip' solution procedure employs a different approach based on series solutions for Papkovich-Neuber potentials \citep{neuber1934neuer}, supplemented with appropriate boundary conditions, namely, zero tangential traction, continuity of normal traction, and continuity of normal displacement along the inclusion-matrix interface \citep{tsuchida1986elastic}.  

Quite notably, despite the differences in interface conditions, all the models (in Fig. \ref{scenario1_fig}) qualitatively predict similar trends and exhibit a similar sensitivity to the stiffness ratio.  
These qualitative trends hold for all stress-free aspect ratios (see Fig. \ref{Appendix:S0} in the Appendix for another example with $\Phi =\Phi_I =2$). From these results, it is already apparent that the isotropic expansion of the biofilm (that is the mode of growth in absence of a confinement) cannot explain the tendency of stiffer inclusions (i.e. inclusions with $\mu_I/\mu_M>>1$) towards a circular state  (Fig. \ref{CaseStudy}). By extending our FEM curves into the deep nonlinear range (see Fig. \ref{Appendix:S0}), we see that the deformed aspect ratio does not continue to decrease towards a circular shape and, in-fact, the trend can reverse for very stiff inclusions. We also note that the QAM results become unreliable in this deeply nonlinear regime. Further,  in this regime, damage must be accounted for.

 
Next, it is instructive to compare the solutions in the two limiting cases of vanishing stiffness ratio and rigid inclusion. In both the limiting cases, we expect the differences due to interface conditions to become negligible and the deformed shapes should unite. Therefore, the FEM solution also serves as a benchmark to compare the accuracy of the QAM in these two limits. Indeed, we see that the QAM results are in reasonable agreement with the FEM predictions for both the limiting cases, despite the simplifying assumptions. In general, we find that the QAM performs  better with increasing stiffness of the inclusion. At the small-strains limit, FEM and LE no-slip results are in agreement, as expected. The QAM and LE with-slip results do not unite even at the small-strains limit. We speculate that this behavior is due to the averaged nature of the incompressibility constraint in the QAM.

 In view of these results, it is worth mentioning the work of \cite{mills2014elastic}, which aims to explain the oblate shapes of  prevascular tumors grown in agarose gel using FEM with a `no slip' condition at the interface. By prescribing similar  isotropic growth (but in 3D) and comparing the elastic energy of the system at $\Delta V/V_0=10$, they conclude that the lowest energy state is the one with the highest aspect ratio (within the considered range of $\Phi=\Phi_I\in[1-3]$). Hence, this study selects among several isotropic expansions the one that has minimal energy (at $\Delta V/V_0=10$) to infer the optimal initial state of the system (at $\Delta V/V_0=0$). This approach implicitly restricts the solution space to isotropic (Scenario \# 1 type) expansions. 
 In contrast,  the current work assumes that the initial state is known and simplifies the deformation field to 2D, but rather than prescribing the growth path, it  explores different growth scenarios and entertains the possibility that the system chooses to evolve its stress-free configuration.   Next, we explore the implications of these different assumptions by examining results for an energy-based  growth law. 



\begin{figure}[h!]
\makebox[\textwidth][c]{\includegraphics[width=0.6\textwidth]{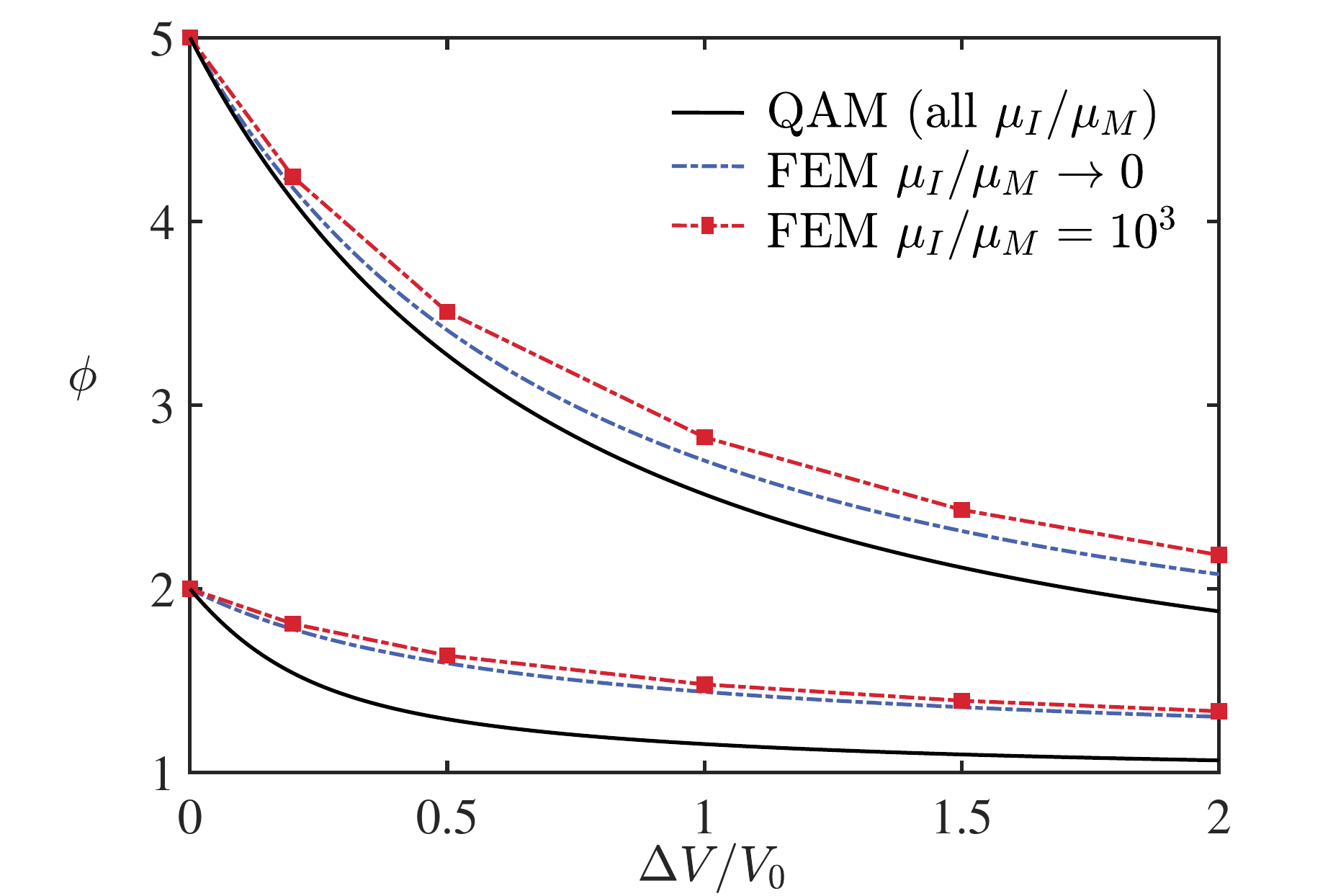}}
\caption{Scenario \#2 results. Deformed shape of the inclusion as a function of the expansion ratio for two different stress-free aspect ratios of the void $\Phi = 2, 5.$  FEM curves show an insensitivity to stiffness ratio. For $\mu_I/\mu_M=10^3$ computation is conducted at discrete points. For the case with vanishing stiffness, computation is conducted by pressurization of the void, hence the curves are continuous.). For the QAM, black curves correspond to all stiffness ratios. }
\label{scenario2_shape_fig}
\end{figure}
\bigskip
\noindent \textbf{Scenario \#2.} In contrast to Scenario \#1, now we allow the inclusion the freedom to choose its stress-free aspect ratio, $\Phi_I$, to minimize the total potential energy of the system at every time. Following the procedure outlined in equations \eqref{axiomA} and \eqref{axiomA_el}, we construct the answer to Questions A and B, to determine the deformed aspect ratio of the inclusion throughout its growth, as shown in Fig. \ref{scenario2_shape_fig}. First, we notice that both FEM and QAM results predict that as the inclusion grows, its deformed shape becomes increasingly circular. By comparison with Fig. \ref{scenario1_fig}, it is  apparent that the undeformed shape is also evolving. Hence, when the inclusion is \textit{active}, that is, it can  adapt  its configuration, the lowest energy state is achieved when its shape becomes more circular with increasing size. In the spirit of D'arcy Thompson's quote cited earlier, rather than growing uniformly to keep the whole shape unchanged, the system  prefers to grow in a non-self-similar way to maintain an energetically favorable configuration. 
Additionally, Fig. \ref{scenario2_shape_fig} shows that both models demonstrate a striking  insensitivity to the stiffness ratio. For the QAM, this insensitivity has been explained analytically, in Section \ref{sec:equil}. Quite interestingly, it appears also for situations in which the two bodies are bonded at the interface (i.e. no slip).  
Note that although results in Fig. \ref{scenario2_shape_fig} are only shown for limiting values of the stiffness ratio, it has been confirmed that this insensitivity occurs in the entire range.
Experimentally, this insensitivity implies that,  in the elastic range,  observation of shape is insufficient to determine the mechanical response of an inclusion. Namely, the fact that it appears to behave  like a fluid, does not imply that it is a fluid.  Instead, subsequent loading of the material system, to deform the matrix and the embedded inclusion can  reveal more information on their respective stiffnesses.

Readers familiar with the theory of classical phase transformations in metallic alloys --- that predicts the formation of spherical precipitates for stiffer inclusions and disk-like precipitates for inclusions softer than the matrix \citep{fratzl1999modeling} --- will notice the apparent discrepancy in the predictions of the two theories.
However, there are a few important differences between the two theories. Firstly, in the case of metallic alloys, a common assumption is to choose the \textit{misfit strain} as a constant that depends only on the lattice parameters of the precipitate and the matrix. By contrast, in the current treatment, the \textit{misfit} keeps increasing with the size of the inclusion and surface energy effects are neglected, thus making the energetic scaling of the two problems different. Secondly, considering the geometric and constitutive nonlinearity of both the matrix and the inclusion, distinguishing between reference and deformed states is of prime importance in the current problem. Lastly, the metallic precipitates, generally, do not have the freedom to evolve their stress-free states as they grow, in sharp contrast to the current growth scenario.

\bigskip 

\noindent \textbf{Scenario \#3.} Here, we examine how damage leads to  change in the deformed shape of the inclusion. At onset, damage may alter the configuration of the matrix resulting in a new observed shape, while the stress-free shape of the inclusion remains the same, as explained in Section \ref{sec:def}.

\begin{figure}[h!]
 \makebox[\textwidth][c]{\includegraphics[width=0.85\textwidth]{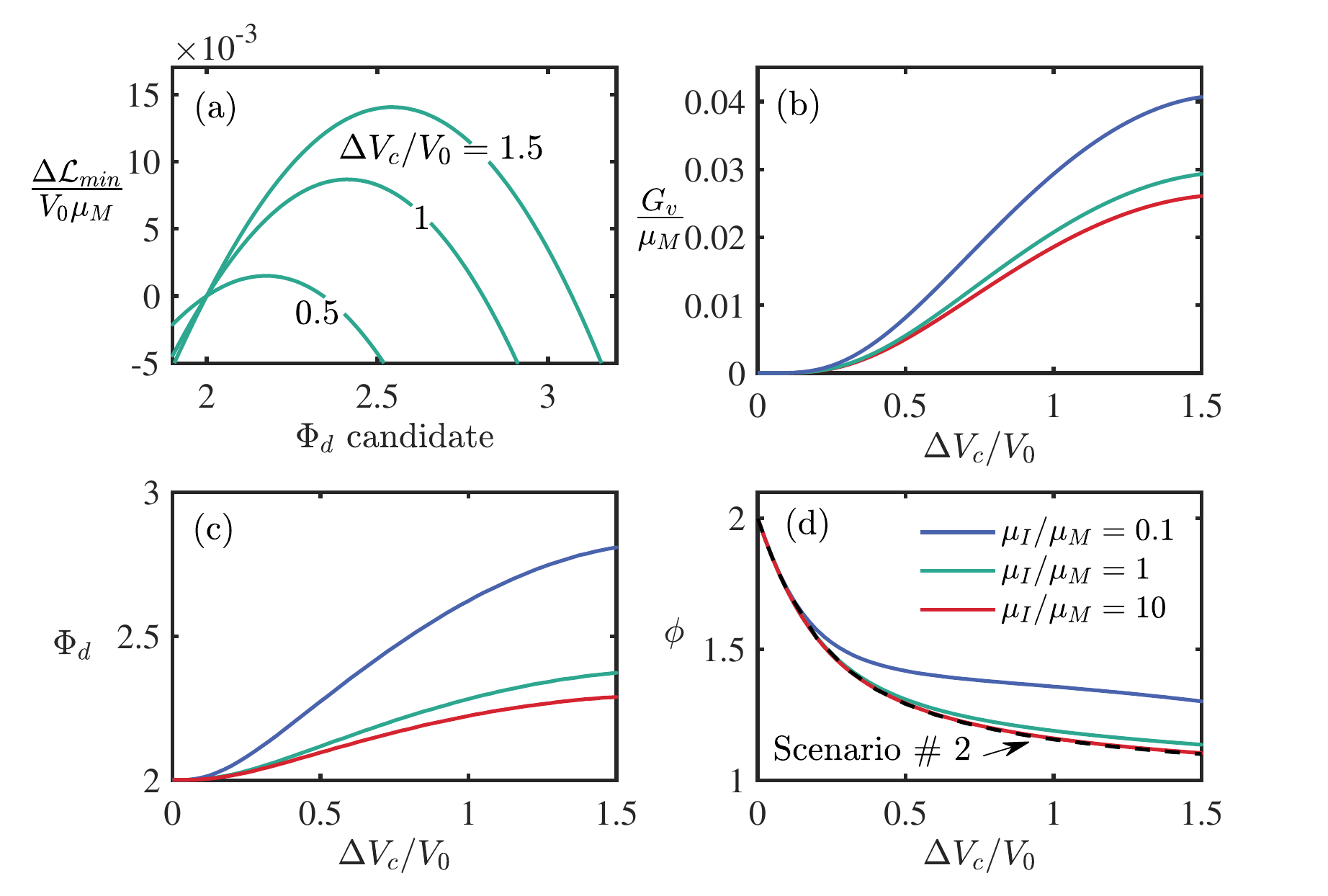}}
\caption{Scenario \#3 results from QAM, indicating the conditions at onset of damage for a given critical expansion $(\Delta V_c)$, from a Scenario \#2 expansion.  (a) Non-dimensionalized $\Delta \mathcal{L}_{min}$as a function of candidate undeformed damaged void aspect ratios for three different critical expansion ratios with $\mu_I/\mu_M =  1$; (b) non-dimensionalized $G_v$ as a function of critical expansion ratio; (c) undeformed aspect ratio of the void at the onset of damage as a function of the critical expansion ratio; (d) deformed shape of the inclusion at the onset of damage as a function of the critical expansion ratio. The black curve represents the path from which damage onsets. }
 \label{scenario3_fig1}
\end{figure}

\noindent\textit{Estimation of $G_v$}: In equation \eqref{Wd} we defined $G_v$ as a material parameter that accounts for energy required per unit volume to alter the configuration of the matrix. Although there are  no direct  methods to  measure the value of this effective parameter, we propose a way to estimate  $G_v$ based on the experimental observations of damage-induced shape change. Assuming that we know the critical volume, $V_c$, at the onset of damage, and the mechanical state of the body (which is fully defined by the pair ($\Phi,\Phi_I$) at $V(t) = V_c$ prior to damage), we  calculate $\mathcal{G}$ from \eqref{G_fun_ellipse} for various candidates of the damaged shape of the void $\Phi_d$.  Finally, from \eqref{damage_cases} we obtain $G_v$ as the maximal value of $\mathcal{G}$ at the onset of damage.

\begin{figure}[h!]
 \makebox[\textwidth][c]{\includegraphics[width=0.85\textwidth]{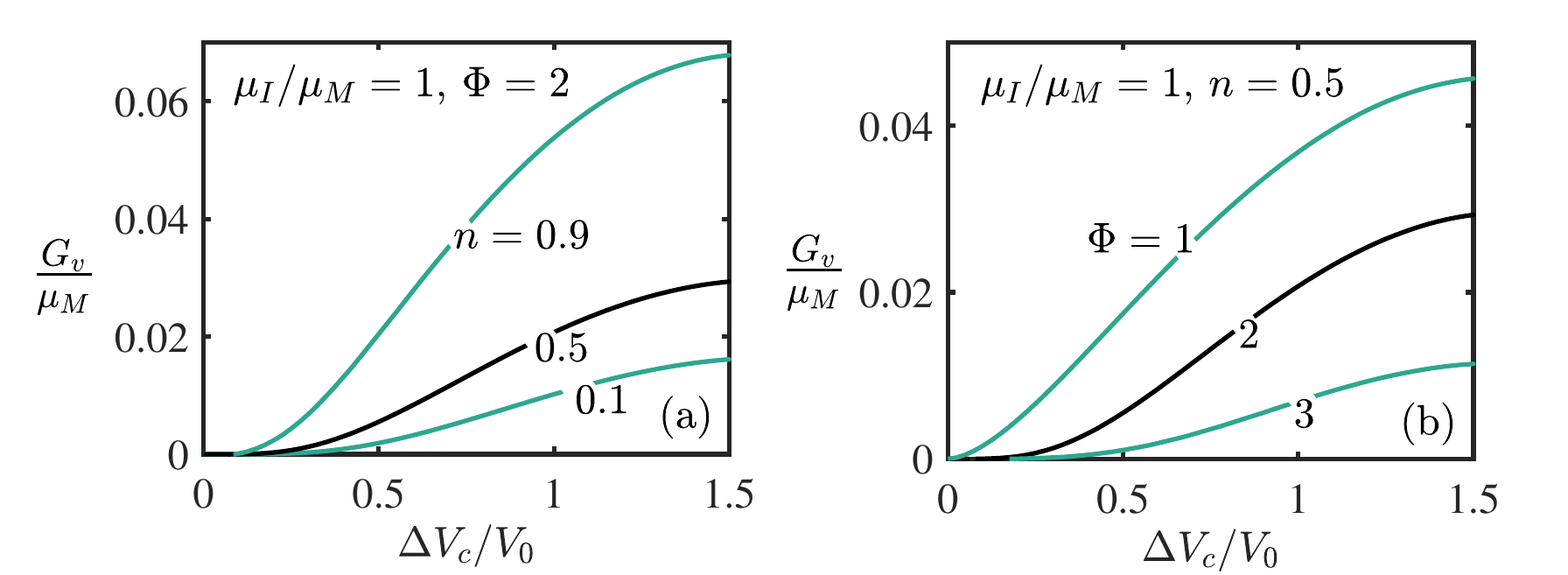}}
 \caption{Sensitivity of $G_v$ to (a) the material parameter $n$ (for $\Phi=2$), and to (b) the void aspect ratio $\Phi$ (for $ n = 0.5$). All curves are shown for $\mu_I/\mu_M = 1$. The black curves are the same in (a) and (b). }
 \label{scenario3_fig2}
\end{figure}

\begin{figure}[h!]
 \makebox[\textwidth][c]{\includegraphics[width=0.85\textwidth]{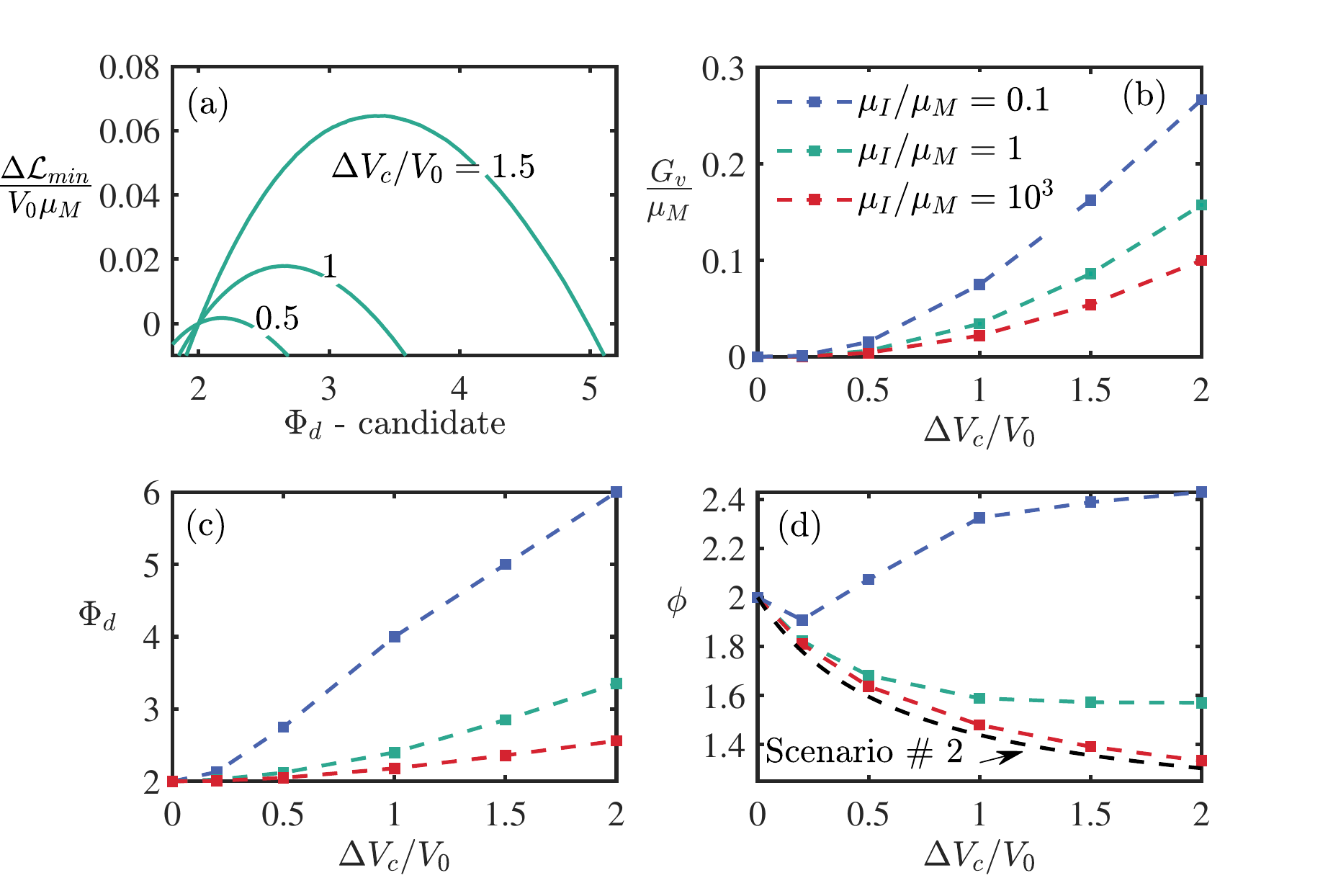}}
 \caption{Scenario \#3 results from FEM, indicating the conditions at onset of damage for a given critical expansion $(\Delta V_c)$, from a Scenario \# 2 expansion.  (a) Non-dimensionalized $\Delta \mathcal{L}_{min}$ for three different critical expansion ratios $\mu_I/\mu_M =  1$; (b) non-dimensionalized $G_v$ as a function of critical expansion ratio; (c) undeformed aspect ratio of the void at the onset of damage as a function of the critical expansion ratio; (d) deformed shape of the inclusion at the onset of damage as a function of the critical expansion ratio. The black curve represents the path from which damage onsets. }
 \label{scenario3_fig3}
\end{figure}

Figs. \ref{scenario3_fig1}-\ref{scenario3_fig3} summarize the results obtained from the above procedure using both the QAM and FEM   for $\Phi= 2,~n=0.5$ and with Scenario \# 2 as the expansion path from which damage initiates. First, for the QAM, the non-dimensionalized plot of $\Delta \mathcal{L}_{min}$ for different values of critical expansion ratio are shown in Fig. \ref{scenario3_fig1}(a) and clearly  exhibit a maximum value that varies in its location with increasing expansion $(\Delta V_c)$. These curves are substituted in \eqref{G_fun_ellipse} to find $G_v={\rm max} \mathcal{G}$ as a function of $\Delta V_c$.  As shown in Fig. \ref{scenario3_fig1}(b), $G_v$ is found to be a monotonic function in this range of the critical expansion ratio, implying  that  if the material has a higher $G_v$, it will damage at a higher expansion ratio. Furthermore, we  observe that for a softer  inclusion to induce damage at the same volume expansion, the matrix must  have a higher $G_v$ (for a given $\mu_M$). Notably, at the onset of damage both the effective undeformed aspect ratio of the void and the deformed aspect ratio always choose a more elongated shape, as seen in Figs. \ref{scenario3_fig1}(c,d), respectively. Nonetheless significant deviations from the Scenario \#2 path upon damage, are only observed for inclusions that are much softer than the matrix. The implications of this result will be further discussed in the context of our observations of biofilm growth.

Next, we examine the sensitivity of the above results to $\Phi$ and the material parameter $n$. Recall that the constitutive parameter $n$ plays a key role in determining the damaged volume \eqref{Vd}. 
For the results shown in Fig. \ref{scenario3_fig2}(a), smaller $n$ results in a higher damaged volume which in turn leads to a smaller value of $G_v$ that is necessary to incur the same energy cost (\ref{Wd}).
Similarly, the stress-free aspect ratio at the onset of damage has a strong bearing on the estimated value of $G_v$ -- more oblate the shape at the onset of damage (comparing at the same critical volume), lower is the $G_v$ (Fig. \ref{scenario3_fig2}(b)).

Finally, we examine the predictions of the FEM for the damage transition in Fig. \ref{scenario3_fig3}. Recall that in contrast to the QAM, here there is perfect bonding between the inclusion and the matrix (i.e. no slip). Again, FEM results are in good qualitative agreement with the predictions from QAM --- $G_v/\mu_M$ increases monotonically with the critical expansion ratio and lower $\mu_I/\mu_M$ implies higher $G_v$. Nonetheless, a more noticeable change of the deformed aspect ratio is found for soft inclusions (Fig. \ref{scenario3_fig3}(d)).  These findings emphasize that differences in interfacial conditions between QAM and FEM do not qualitatively change the behavior of the system. Similar behaviors appear also for different initial aspect ratios ($\Phi$). It should be noted however that we restrict our investigation to moderate expansions where the model assumptions apply.

\bigskip

\begin{figure}[h!]
 \makebox[\textwidth][c]{\includegraphics[width=0.9\textwidth]{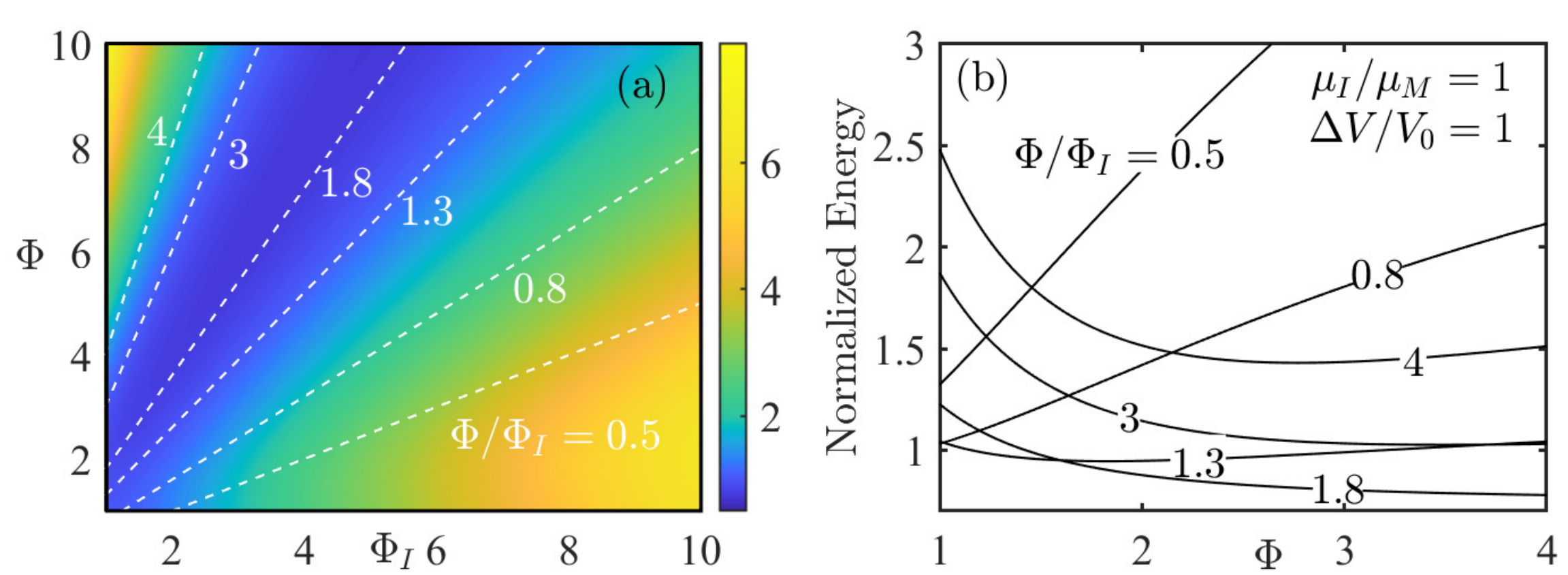}}
 \caption{Energy landscape for scenario \#4 (with $\Delta V/V_0 = 1$ and $\mu_I/\mu_M=1$). (a) Colormap showing the energy landscape, where the dashed straight lines represent constant values of $\Phi/\Phi_I$; (b) elastic energy as a function of $\Phi$, along different lines of constant $\Phi/\Phi_I$. In both plots the energy is normalized by $\mu_MV_0$.}
 \label{heatmap}
\end{figure}

\noindent \textbf{Scenario \#4.} In this scenario both the inclusion and the matrix can remodel themselves to minimize the total energy. Thus, the system has increased freedom when compared to Scenario \#2 and, as illustrated in Fig. \ref{fig:S4}, an entire 2D plane of solutions for $(\Phi,\Phi_I)\in[1,\infty)$ must be scanned to identify optimal solutions. Given the long computation times that this would entail using the FEM, and having established that the results of the FEM and the QAM lead to qualitatively  similar results, we take advantage of the simplicity of QAM to determine the growth path.  
Fig. \ref{heatmap}(a) shows  the energy landscape in the $(\Phi,\Phi_I)$ plane for $\Delta V/V_0 =1$ and with $\mu_I/\mu_M =1$. Different shades imply different normalized energy and the various lines indicate fixed ratios $\Phi/\Phi_I$. A valley is observed near $\Phi/\Phi_I=1.8$. To determine where the minimum is found along this valley, we examine   the normalized energy as a function of $\Phi$ along the lines plotted of constant $\Phi/\Phi_I$ in  Fig. \ref{heatmap}(b). It is clearly shows that the energy monotonically  decrease along the $\Phi/\Phi_I=1.8$. Hence, the system will choose a `sliver-like'  shape with infinite aspect ratio. We find that these results hold for any stiffness ratio. Therefore, when the system has the freedom to remodel both the inclusion and the matrix, it is always energetically favorable to expand along the major axis and become more oblate\footnote{Although not a complete investigation of the energy landscape, anecdotal results obtained from the FEM confirm this trend also for the case with no slip.}. Clearly, if surface tension effects exist,  this tendency would be restricted. 

\noindent \textbf{Analysis of Bacterial Biofilms.} Equipped with the knowledge of different scenarios, we can now return to our case study (Section \ref{sec:CS}) in an attempt to elucidate the morphological evolution of bacterial biofilms. Initially, as the bacteria start to multiply and the biofilm grows, it is anticipated that elastic deformation is the dominant mode in the matrix, and thus that the system undergoes Scenario \#2 type expansion wherein the biofilm inclusion has the freedom to rearrange internally, which in absence of any other apparent physical mechanisms that can drive changes of shape,  allows it to achieve the state with lowest total energy. Recall that Scenario \#2 expansion is insensitive to the respective stiffnesses of the matrix and the inclusion.  Indeed, in our experiments, tracking of several colonies did not reveal any clear correlation between the growth path and the stiffnesses, thus providing further evidence to support the hypothesis of an energy-based growth law. However, it is not possible to make a strong conclusion based on the data in this regard, due to the  uncertainty in quantifying the initial conditions of the biofilm-agarose system.

\begin{figure}[h!]
 \makebox[\textwidth][c]{\includegraphics[width=0.9\textwidth]{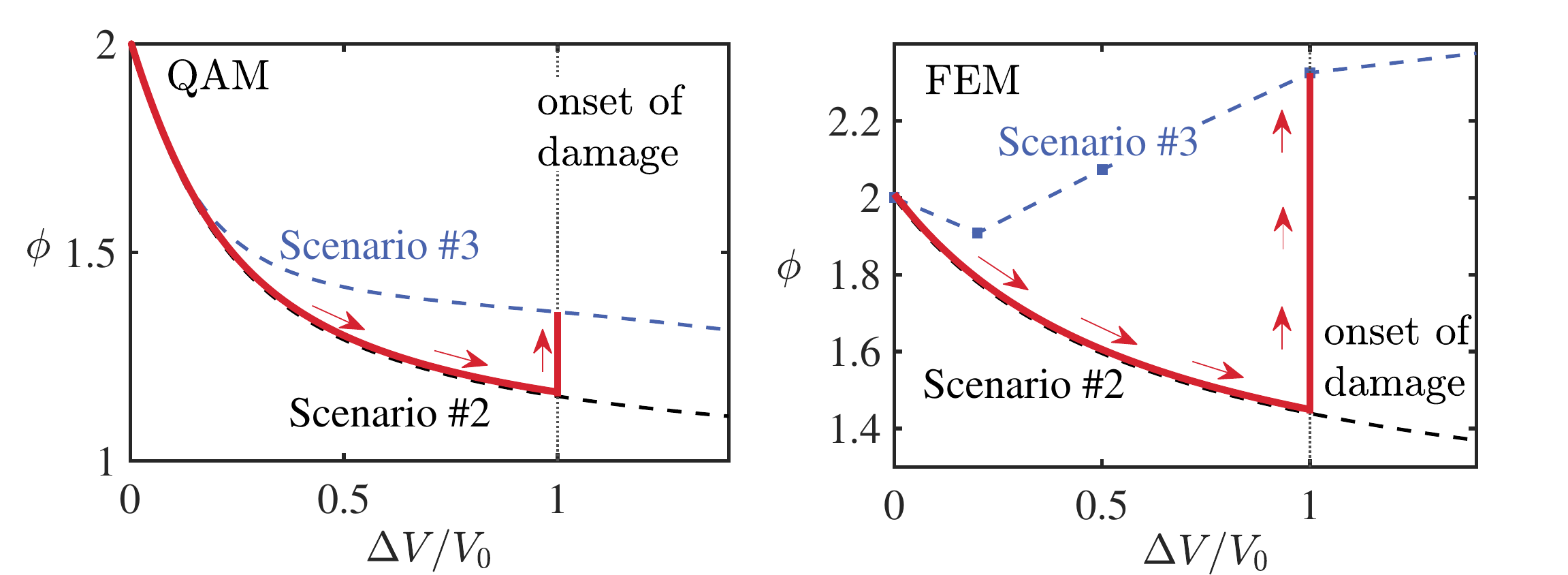}}
 \caption{Model-based explanation of the shape evolution of bacterial biofilms. The red curve represents the typical  path of the system  with a softer inclusion (shown for $\Phi=2$ with $\mu_I/\mu_M = 0.1$); following  a Scenario \#2 type expansion, at onset of damage, transition to a more elongated shape (via Scenario \#3) is shown  using both QAM and FEM. For stiffer inclusions ($\mu_I/\mu_M > 1$), Scenarios \#2 and \#3 curves become indistinguishable and thus onset of damage does not induce an apparent change of shape.}
 \label{fig:CS_results}
\end{figure}

It is conceivable that morphogenesis along the Scenario \#2 branch is limited by the onset of damage in the embedding media (agarose).  Hence, after monotonically tending towards a spherical shape, at a critical threshold (defined by the yield stress/strain of the confining gel), the agarose gel starts to accumulate diffuse damage that affects the observed shapes of the biofilm, but without fracturing it. Accordingly, we expect that at the onset of damage, the system undergoes a Scenario \#3 type transition whereby the matrix can instantaneously change its stress-free shape. In making this transition, the stiffness ratio between the biofilm and the gel plays a crucial role, as shown in Figs \ref{scenario3_fig1} and \ref{scenario3_fig3}. In particular, it is found that if the biofilm is softer, the configuration becomes more oblate, as shown through an example morphogenesis pathway  for $\mu_I/\mu_M = 0.1$ in Fig. \ref{fig:CS_results}. In contrast, if the biofilm is stiffer, the damage branch is indistinguishable from Scenario \#2 and the biofilm  retains its deformed aspect ratio.  Together, the soft and stiff cases explain the two different trends observed in our Case Study (Fig. \ref{CaseStudy}). While our purely mechanical model predicts a sudden jump in state upon damage, in real systems such as bacterial biofilms, this transition will manifest slowly due to rate dependent effects arising from the viscoelastic nature of both the biofilm and the agarose gel.

Furthermore, by studying the two extreme interfacial conditions between the inclusion and the matrix and obtaining very similar qualitative trends, we can confidently remark that the current theory presents a minimal model that explains the major morphological trends. A more refined picture for the interfacial conditions in bacterial biofilms and other biological and engineering systems that covers the spectrum between no-slip and perfectly bonded interface, and also accounts for surface energy considerations is an important future step in the modeling of these systems.

\section{Concluding remarks}

Instances of growth and morphogenesis of bodies, or inclusions, as they are embedded in a confining medium, are ubiquitous in both natural and engineered systems, with examples ranging from the growing roots of a plant, to the formation of precipitates in metals. However, how the accumulating mechanical stresses and the driving mechanisms of growth collaborate to determine the morphological fate of the growing body is not well understood. In this work, we show that the mechanical confinement can have a nontrivial effect on the  morphogenesis. To capture the ability of the material  to adapt its configuration by way of internal reorganization or damage, we consider various growth scenarios by which the system can choose to evolve  as it deforms into the nonlinear range. Motivated by our experimental case study that examines the growth of confined biofilms, we put a special emphasis on determining a natural growth law that dominates in absence of an inherent tendency towards a specific  configuration. Hence, we postulate that morphological evolution will proceed to minimize the total energy of the system. Further, to model realistic growth scenarios  we extend Eshelby's classical inclusion theory into the nonlinear range by developing  minimal quasi-analytical and  computational (finite element) approaches. We limit our attention to incompressible bodies undergoing plane-strain deformation, and consider two physical limits of the interfacial condition spectrum (perfect bonding and perfect slip). 
Advances on these two fronts allow us to mathematically pose and answer fundamental questions concerning the evolving configuration and the observed shape of the growing inclusion, under different representative scenarios. We apply this development to explain the morphological evolution of bacterial biofilms including the intriguing non-monotonic transition from a tendency towards a circular shape to the subsequent  elongation towards an oblate shape. Our modeling reveals that for the class of elliptic cylindrical inclusions, the energy-based growth law predicts insensitivity to the respective stiffnesses of the inclusion and the matrix. Furthermore, diffuse damage in the matrix that may emerge at large volumetric expansions,  drives the shape  away from circular to elongated if the inclusion is softer than the matrix. 

While the current work addresses the coupling between the fate of a growing body and its mechanical confinement, it is a first step towards gaining a more comprehensive understanding of the role of environmental factors in growth of a complex body -- both geometry and material-wise.
A natural extension  is towards the modeling of three-dimensional inclusions. Furthermore, a variety of interfacial considerations, such as partially slipping interactions, surface energy, dynamically evolving interfaces, and frictional slipping, are also opportunities to extend the modeling framework. Beyond the onset of damage, future work should center on examining the incremental evolution of damage and the potential occurrence of  fracture that may dominate at larger volume expansions. Finally, it remains to extend this framework to more complex growth scenarios in which an inherent developmental blueprint is confronted by an external constraint, much like an embryo growing inside the womb, or fruit growing in confinement (Fig. \ref{tomato}).

\section*{Acknowledgements}
We thank Hannah Varner for helpful discussions. T.C. acknowledges the support of Dr. Timothy B. Bentley, Office of Naval Research Program Manager, under award number N00014-20-1-2561, and support from the National Science Foundation under award number 1942016. J.Y. holds a Career Award at the Scientific Interface from the Burroughs Welcome Fund. 


\section*{Appendix}

\setcounter{figure}{0}
\renewcommand{\thefigure}{A\arabic{figure}}

\begin{figure}[h!]
  \centering
    \includegraphics[width=0.9\textwidth]{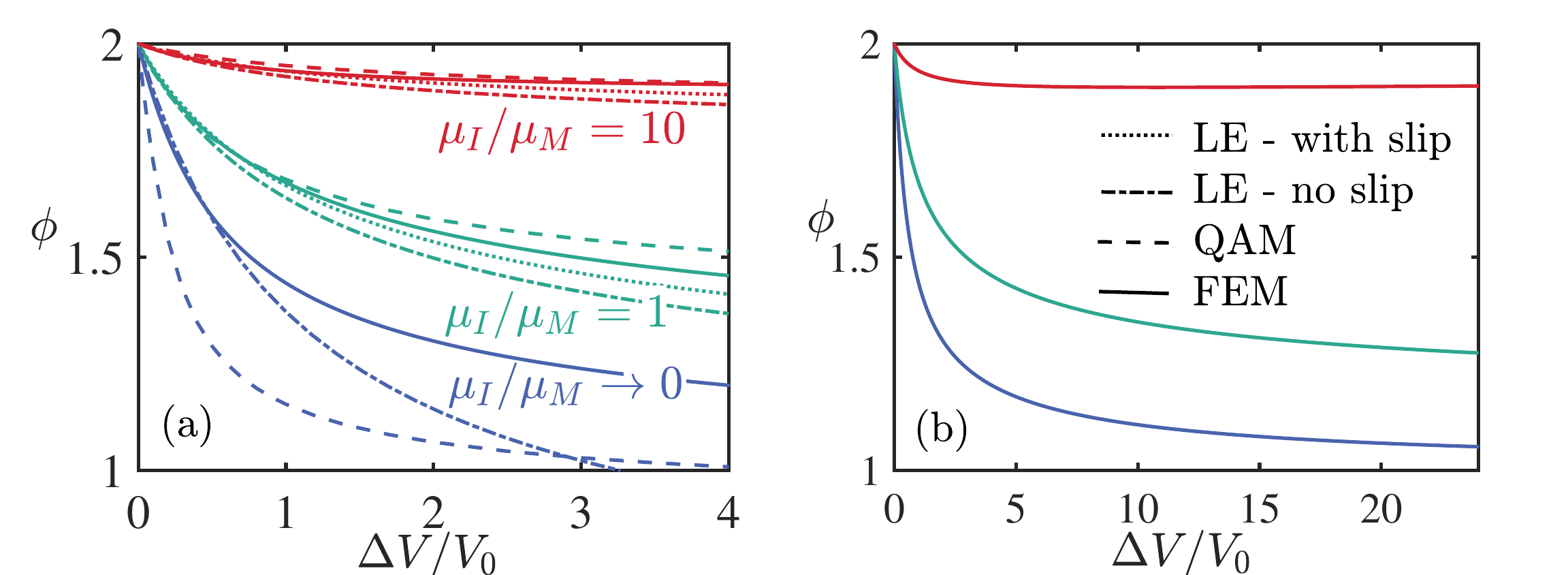}
  \caption{Scenario \#1 results for isotropic expansion with  ($\Phi=\Phi_I = 2$). On the left - the deformed shape of the inclusion is shown as a function of the expansion ratio using for different models that account for the two limits of the interface condition. On the right - we extend only the FEM curves to larger expansion, showing saturation of the aspect ration, and even a change in trend for $\mu_I/\mu_M=10$.}  \label{Appendix:S0}
\end{figure}

\bibliographystyle{elsarticle-harv}
\bibliography{refs} 
\end{document}